# Ligation of random oligomers leads to emergence of autocatalytic sequence network

Patrick W. Kudella[1], Alexei V. Tkachenko[2], Sergei Maslov[3,4], Dieter Braun*[1]

[1]Systems Biophysics and Center for NanoScience, Ludwigs-Maximilian-Universität München, 80799 Munich, Germany
[2]Center for Functional Nanomaterials, Brookhaven National Laboratory, Upton, New York 11973, USA
[3]Department of Bioengineering, University of Illinois at Urbana-Champaign, Urbana, Illinois 61801, USA
[4]Carl R. Woese Institute for Genomic Biology, University of Illinois at Urbana-Champaign, 1206 West Gregory Drive, Urbana Illinois 61801, USA

## ABSTRACT

The emergence of longer information-carrying and functional nucleotide polymers from random short strands was a major stepping stone at the dawn of life. But the formation of those polymers under temperature oscillation required some form of selection. A plausible mechanism is template-based ligation where theoretical work already suggested a reduction in information entropy.

Here, we show how nontrivial sequence patterns emerge in a system of random 12mer DNA sequences subject to enzyme-based templated ligation reaction and temperature cycling. The strands acted both as a template and substrates of the reaction and thereby formed longer oligomers. The selection for templating sequences leads to the development of a multiscale ligation landscape. A position-dependent sequence pattern emerged with a segregation into mutually complementary pools of A-rich and T-rich sequences. Even without selection for function, the base pairing of DNA with ligation showed a dynamics resembling Darwinian evolution.

## BACKGROUND

One of the dominant hypotheses to explain the origin of life[1–3] is the concept of RNA world. It is built on the fact that catalytically active RNA molecules can enzymatically promote their own replication[4–6] via active sites in their three dimensional structures[7–9]. These so-called ribozymes have a minimal length of 30 to 41 bp[9,10] and, thus, a sequence space of more than $4^{30} \approx 10^{18}$. The subset of functional, catalytically active sequences in this vast sequence space is vanishingly small[11] making spontaneous assembly of ribozymes from monomers or oligomers all but impossible. Therefore, prebiotic evolution has likely provided some form of selection guiding single nucleotides to form functional sequences and thereby lowering the sequence entropy of this system.

The problem of non-enzymatic formation of single base nucleotides and short oligomers in settings reminiscent of the primordial soup has been studied before[12–16]. However, the continuation of this evolutionary path towards early replication networks would require a pre-selection mechanism of oligonucleotides (as shown in Fig. 1a), lowering the information entropy of the resulting sequence pool[17–20]. In principle, such selection modes include optimization for information storage, local oligomer enrichment e.g. in hydrogels or in catalytically functional sites.

An important aspect of a selection mechanism is its non-equilibrium driving force. Today's highly evolved cells function through multistep and multicomponent metabolic pathways like glycolysis in the Warburg effect[21] or by specialized enzymes like ATP synthase which provide energy-rich adenosine triphosphate (ATP)[22]. In contrast, it is widely assumed[3,4,23–26] that selection mechanisms for molecular evolution at the dawn of life must have been much simpler, e.g. mediated by random binding between biomolecules subject to non-equilibrium driving forces such as fluid flow and cyclic changes in temperature.

Here, we explored the possibility of a significant reduction of sequence entropy driven by templated ligation[17] and mediated by Watson-Crick base pairing[27]. Starting from a random pool of oligonucleotides we observed a gradual formation of longer chains showing reproducible sequence landscape inhibiting self-folding and promoting templated ligation. Here we argue, that base pairing combined with ligation chemistry, can trigger processes that have many features of the Darwinian evolution.

As a model oligomer we decided to use DNA instead of RNA since the focus of our study is on base pairing which is very similar for both[28]. We start our experiments with a random pool of 12mers formed of bases A (adenosine) and T (thymine). This binary code facilitates binding between molecules and allows us to sample the whole sequence space in microliter volumes ($2^{12} << 10\ \mu M * 20\ \mu l = 10^{14}$).

Formation of progressively longer oligomers from shorter ones requires ligation reactions, a method commonly employed in hairpin-mediated RNA and DNA replication[29,30]. At the origin of life, this might have been achieved by activated oligomers[31,32] or activation agents[33–35]. Our study is focused on inherent properties of self-assembly by base pairing in random pools of oligomers and not on chemical mechanisms of ligation. Hence, we decided to use TAQ DNA ligase - an evolved enzyme for templated ligation of DNA[19]. This allowed for fast turnovers of ligation and enabled the observation of sequence dynamics.

## RESULTS

To test templated elongation of polymers in pools of random sequence oligomers, we prepared a 10 µM solution of 12mer DNA strands composed of nucleotides A and T (sequence space: 4096) and subjected it to temperature cycling, similar to reference[19] with 20 s at denaturation temperature of 75 °C and 120 s at ligation temperature of 33 °C. Temperatures were selected according to the melting dynamics of the DNA pool; the time steps were prolonged relative to Toyabe and Braun (SI section 5.3) because of a greater sequence space. The sample was split into multiple tubes and exposed to 200, 400, 600, 800, 1000 temperature cycles, with one tube kept at 4 °C for reference.

To study the length distributions in our samples we used polyacrylamide gel electrophoresis (PAGE, Fig. 1d). The first lane is the reference sequence not exposed to temperature cycling, where small amounts of impurities are visible at short lengths (SI Section 3.1). The latter lanes show the temperature-cycled samples. As the number of cycles increases, progressively longer strands in multiples of 12 emerge, as the original pool only consisted of 12mers. Fig. 1c shows the concentration quantification of each lane (compare SI section 3). For higher cycle counts the total amount of products increases and the concentration as a function of length decreases slower. The behavior of this system is dependent on the time and temperature for both steps in the temperature cycle, the monomer-pool concentration and the sequence space of the pool (SI section 5).

An important property of the initial monomer-pool is its sequence content. Although for pools with lower sequence complexity it is possible to show different strand compositions using PAGE[36,37], a large size of our "monomer" ($2^{12}$=4096) and 24mer product pools (sequence space: $2^{24} \approx 16.8 \times 10^6$) excludes this approach. Thus, we analyzed our final products by Next Generation Sequencing (NGS) to get insights into product strand compositions.

Plotting the probability of finding a base at a certain position (Fig. 1c inset) revealed no distinct pattern in 12mers other than a slight bias towards As. However, longer chains starting with 24mers developed a strikingly inhomogeneous sequence pattern: bases around ligation sites show a distinct AT-alternating pattern, while regions in the middle of individual 12mers are preferentially enriched with As.

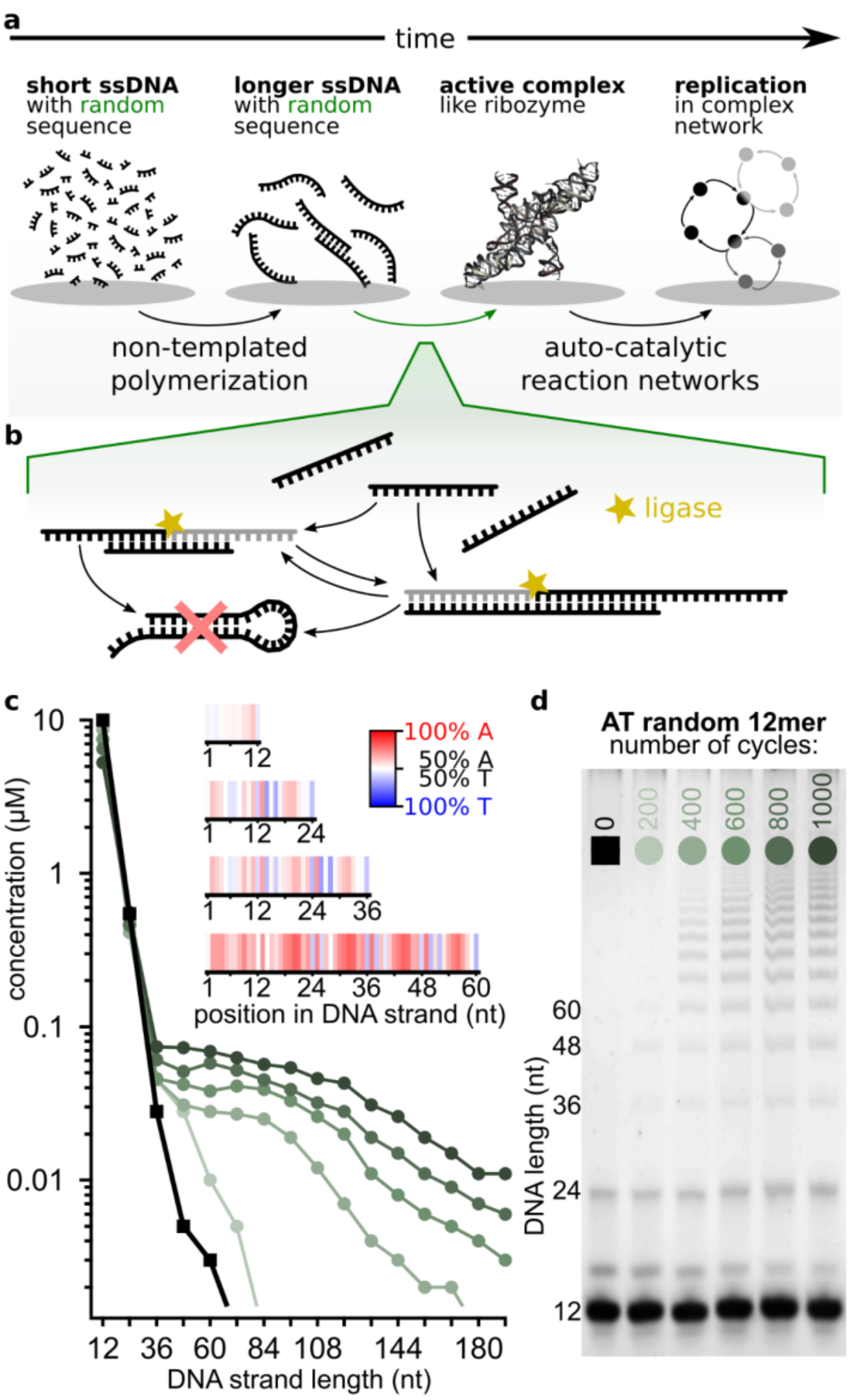

***Fig. 1, Autocatalytic templated ligation of DNA 12mers.***
***a** Before cells evolved, the first ribozymes were thought to perform basic cell functions. In the exponentially vast sequence space, spontaneous emergence of a functional ribozyme is highly unlikely, therefore pre-selection mechanisms were likely necessary.*
***b** In our experiment, DNA strands hybridize at low temperatures to form 3D complexes which can be ligated and preserved in the high temperature dissociation steps. The system self-selects for sequences with specific ligation site motifs as well as for strands that continue acting as templates. Hairpin sequences are therefore suppressed.*
***c** Concentration analysis shows progressively longer strands emerging after multiple temperature cycles. The inset (A-red, T-blue) shows that while 12mers (88009 strands) have essentially random sequences (white), various sequence patterns emerge in longer strands (60mers, 235913 strands analyzed).*
***d** Samples subjected to different number (0-1000) of temperature cycles between 75 °C and 33 °C. Concentration quantification is done on PAGE with SYBR post-stained DNA.*

The information entropy of longer chains is expected to be smaller than the entropy of a random sequence strand of the same length, if some sort of selection mechanism is involved[17]. We analyzed the entropy reduction for different lengths of products (Fig. 2a) as well as the positional dependence of the single base entropy for 60mer products (Fig. 2b). The relative entropy reduction is similar to one used in Derr *et. al*[38] where 1 describes a completely random ensemble and 0 an ensemble of only one sequence. Entropy reduction was observed in all analyzed product lengths with a greater reduction observed for longer oligomer lengths. The entropy of each 12mer subsequence was also found to be significantly lower than that of random 12mers (Fig. 2b, black line). The central subsequence had the lowest entropy while 12mers located at both ends of chains had relatively higher entropies. This behavior was also observed as a function of nucleotide position within a 12mer suggesting a multi-scale pattern of entropy reduction.

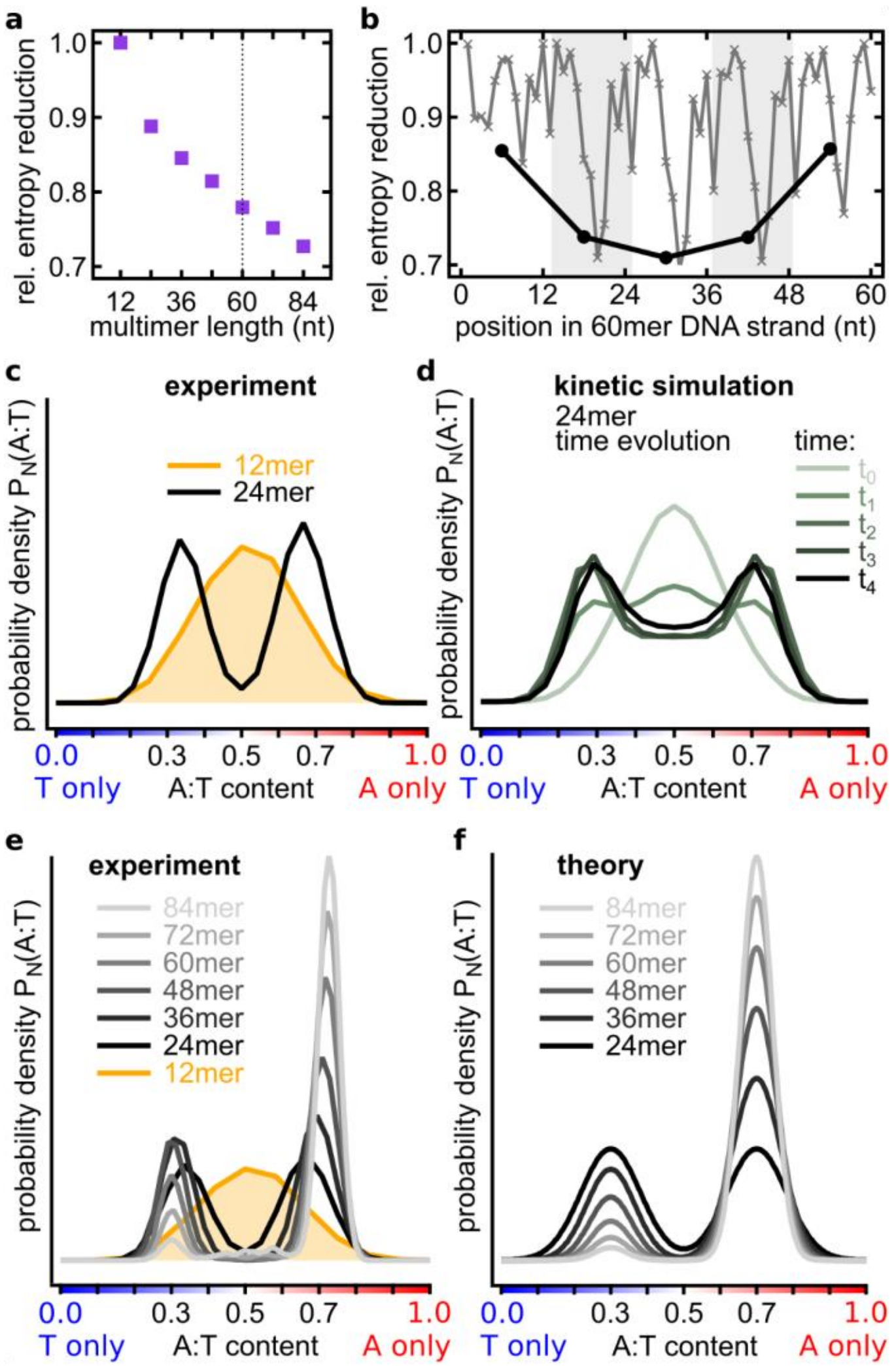

***Fig. 2, Hairpin formation amplifies selection into A-rich and T-rich sequences.***
***a*** *Relative entropy reduction as a function of multimer product length: 1 – a random pool and 0 – a unique sequence.*
***b*** *Relative entropy reduction of 60mer products. Black: Entropy reduction of 12 nt subsequences compared to a random sequence strand of the same length. Grey: Entropy reduction at each nucleotide position showing positional dependence.*
***c*** *A gradual development of the bimodal distribution of A:T ratio in chains of different lengths. While the A:T ratio in 12mers has a single-peaked nearly binomial distribution, 24mers already have a clearly bimodal distribution peaked at 65:35 % (A-type strands) and 35:65 % (T-type strands) A:T ratios.*
***d*** *Emergence of a bimodal distribution in a kinetic model of templated ligation. Sequences with nearly balanced A:T ratios are prone to formation of hairpins. In the model these hairpins prevent strands from acting as templates and substrates for ligation reactions thereby suppressing the central part of the distribution.*
***e*** *A:T ratio distributions in strands of different length. As length increases A-type strands become progressively more abundant in comparison to T-type strands.*
***f*** *A:T ratio distributions in a phenomenological model taking into account a slight AT-bias in the initial 12mer pool resemble experimentally measured ones (panel e).*

In the initial pool of random 12mers the A-to-T ratio distribution is shaped binomially, as expected for a random distribution. However, it dramatically shifted for 24mer products of ligation: a bimodal distribution of about 65:35 % A:T (A-type) as well as the inverse, 35:65 % A:T (T-type) was observed with 24mer products (Fig. 2c). DNA strands composed of only two complementary bases are more prone to formation of single-strand secondary structures like hairpins than DNAs composed of all four bases. In our templated ligation reaction, we expected that hairpin-sequences are not elongated and also not used as template-strands because they form catalytically passive Watson-Crick-base-paired configuration. A bimodal AT-ratio distribution (Fig. 2d) also emerged in a kinetic computational model in which a pool of random 12mers was seeded with a small initial amount of random sequence 24mers. 24mers that formed hairpins could not act as templates and were therefore less likely to be reproduced (see SI for details of this model, section 18.2).

For longer products the bimodal distribution got sharper and centered at approximately 70:30 % A:T and 30:70 % A:T (Fig. 2e). To compare the distributions of different lengths we computed probability density functions (PDF) of A:T fractions. Each distribution is the sum (integral) over all probabilities $P_N$ to find a certain A:T-fraction $\mathrm{d}_{A:T}$ in chains of length $N$:

$$\int P_N(A{:}T)\mathrm{d}_{A:T} = 1. \quad (1)$$

The main difference of longer oligomers was a rapid increase of the ratio between the number of A-type and T-type sequences. As oligomers get longer the effect becomes more pronounced. This might be a result of a small bias in the initial pool which has slightly more monomers of A-type than T-type (SI section 9.1).

As predicted theoretically[39], the eventual length distribution is approximately exponential. A small A-T bias leads to the respective average chain lengths, $\bar{N}_A$ and $\bar{N}_T$, to be somewhat different for the two subpopulations. As a result, the bias gets strongly amplified with increasing chain length:

$$P_N(A{:}T) \sim \exp\left(-N\left(\frac{1}{\bar{N}_A} - \frac{1}{\bar{N}_T}\right)\right) = \beta^{-N/12}. \quad (2)$$

A simple phenomenological model can successfully capture the major features of the observed A:T PDFs for multiple chain lengths. Specifically, we assume both A-type and T-type sub-populations -to maximize the sequence entropy, subject to the constraint that the average A:T content is shifted from the midpoint (50:50 % composition), by values $\pm x_0$, respectively. This model presented in SI section 18.1, results in a distribution that strongly resembles experimental data, as shown in Fig. 2e-f A:T profiles for all chain length are fully parameterized by only two fitting parameters: $\beta = 0.785$, and $x_0 = 0.2$.

The proposed mechanism of selection of A-type and T-type subpopulations due to hairpin suppression is further supported by direct sequence analysis. Fig. 3b shows PDFs of the longest sequence motifs that would allow hairpin-formation, across the entire pool of sequences of given lengths. While the overall chain length increased by a factor of seven (12 to 84 nt), the most likely hairpin length only grew by a factor of 1.89 (3.7 to 7 nt) (Fig. 3b). The observed relationship between the strand length $N$ and the most likely hairpin length $l_0$ can be successfully described by a simple relationship obtained within the above described maximum-entropy model. Specifically, for a random sequence with bias parameter $p = 0.5 + x_0$, one expects $N$ to be related to $l_0$ as follows (as in Fig. 2f):

$$N = 2l_0 + \sqrt{2}(2p(1-p))^{-l_0/2}. \quad (3)$$

As one can see in Fig. 3c, this result is in an excellent agreement with experimental data for all the long chains, assuming $p$=0.785. This A:T ratio is indeed comparable to the one observed in the A-type subpopulation. On the other hand, the maximum probability length of the longest hairpin for 12mers is consistent with an unbiased composition, $p$=0.5.

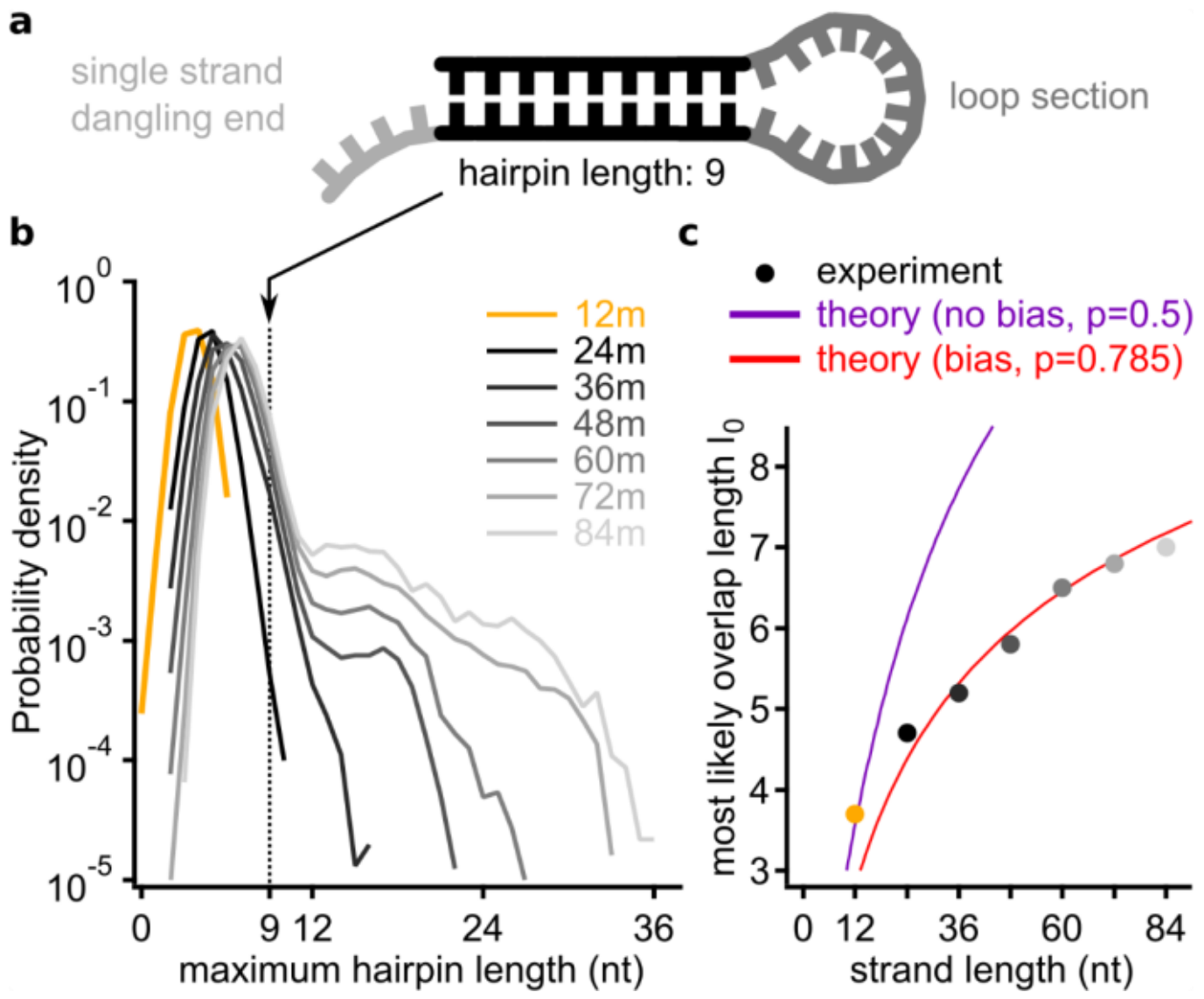

***Fig. 3, Large scale entropy reduction and sequence correlation per strand.***
***a*** *Sketch of a single strand DNA secondary structure folding on itself, called hairpin. The double stranded part is very similar to a standard duplex DNA.*
***b*** *Comparing the PDFs of the maximum hairpin length for all strands reveals a group of peaks at around 4 to 7 nt, increasing with the DNA length. Starting for 48mers, there is a tail visible: these self-similar strands are more abundant, the longer the product grows (compare A:T fraction close to p=0.5 in Fig. 2c).*
***c*** *The peak-positions as function of the product length follow equation (3). The unbiased 12mers are on the curve with coefficient p=0.5, whereas the products starting from 36mers lay on the curve with p=0.785. The bias parameter p is derived from the PDFs in Fig. 2d and describes the A:T-ratio in the strand.*

While hairpin formation inhibits the self-reproduction based on template-based ligation, Fig. 3b reveals another dramatic feature: a small fraction of chains does feature very long hairpin-forming motifs (seen as shoulders in the distribution function). This effect also reveals itself as small peaks on the 84mer curve in Fig. 2e. Those peaks around A:T ratio 0.4, 0.5 and 0.6. stem from subpopulations that have multiple A-types as well as multiple T-type subsequences (see SI section 12) and are prone to hairpin formation.

The mechanism of formation of these self-binding sequences may involve recombination of shorter A-type and T-type chains, or self-elongation of shorter hairpins. In either case, the harpin sequence cannot efficiently reproduce by means of template ligation. However, the reminder of the pool would keep producing them as byproduct. Ironically, for the templated ligation reaction this is a possible failure mode, but those long hairpins may play a key role in the context of origin of life, as precursors of functional motifs. For instance, work by Bartel and Szostak[11,40] identifies RNA self-binding as crucial for the direct search of ribozymes – those molecules need to fold into non-trivial secondary structures to gain their catalytic function.

The separation into A-type and T-type subpopulations only accounts for a small part of the sequence entropy reduction. The emerging ligation landscape in the sequence space is far richer.

Sequence analysis of the junctions in-between original 12mer revealed additional information about that landscape, already hinted by patterns seen in Fig. 1b. We characterize pairs of junction-forming sequences with their Z-scores, i.e. probability of their occurrence scaled with its expected value and divided by the standard deviation calculated in the random binding model (see SI section 14).

Fig. 4a shows Z-score heatmaps for junctions within A-type (left panel) and T-type (right panel) subpopulations. More specifically, we show sequences left (row) and right (column) of the junction between the 4th to the 5th 12mers in the respective 72mer. These heatmaps reveal a complex landscape of over- and under-represented junction motifs shown respectively in dark-teal and dark-ocher colors. Emergence of such complex landscape has been theoretically predicted in Ref. [17] Landscape peaks include repeating A-T motif of alternating bases crossing the ligation site (dark-teal peak near the center of each of both heatmaps). Relatively rare motifs (valleys) correspond to poly-A and poly-T sequences extending across the junction (dark-ocher areas). One exception to this rule is a relatively abundant poly-A motif at the bottom right of the A-type heatmap (light-teal). Interestingly, these junction sequences had AT-patterns in the beginning of the "left side" and the end of the "right side". This might provide a clue to the origin of these "abnormal" junction motifs. Indeed, they may have been templated by abundant poly-T sequences in the middle of T-type 12mers flanked by alternating A-T motifs. In other words, junctions at templates of poly-A junction motifs may have been shifted by 6 nt relative to substrates. Actually, substrates have no restriction on where they may hybridize on a long template and might happen to have their ligation site in the region of poly-T of the template strand. We call this "ligation site shift", as explained in SI section 16. Other preferred junction subsequences include repetitions of the AAT motif across the junction (the dark-teal peak in the upper left corner of the left panel).

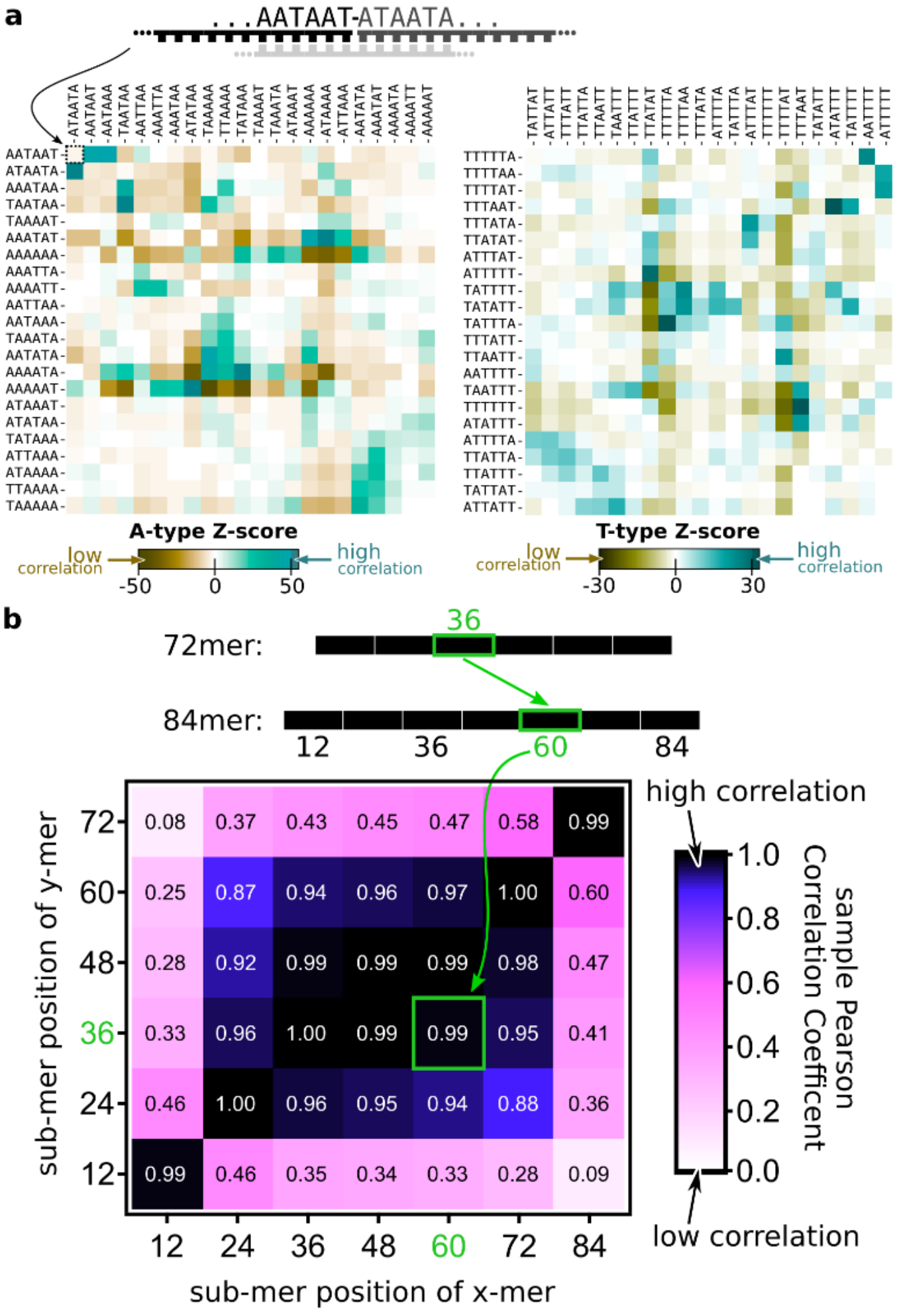

***Fig. 4, Emergent landscape of junction sequences.***
***a** The heatmap of Z-scores quantifying the probability to find a junction between a 6 nt sequence listed in rows followed by the 6 nt sequence listed in columns compared to finding it by pure chance and normalized by the standard deviation. Z-scores were calculated for the junction between 4th to the 5th 12mers in 72mers of A-type (left) and T-type (right) respectively. Other internal junctions in all long chains form very similar landscapes composed of over- (teal) and under-represented (ocher) sequences and described in detail in the text. T-type sequences complementary to A-type sequences correspond to the 90° clockwise rotation of the left panel (note a similarity of landscapes in two panels after this transformation).*

***b*** *The matrix of sample Pearson Correlation coefficients between abundances of 12mers in different positions (1 to 6) inside 72mers (rows) and 84mers (columns). Light regions mark low correlations, dark regions mark high correlations. Very high correlations (>0.9) at the center of the table mean that very similar sequences get selected at all internal positions of chains of different lengths. Different selection pressures operate on the first 12mer and the last 12mer of a chain, yet their sequences are similar in chains of different lengths.*

How similar are selective pressures operating on sequences of different 12mers within longer chains? Fig. 4b quantifies this similarity in terms of sample Pearson-Correlation-Coefficient (sPCC) between abundances of 12mer sequences in different positions of long chains of different lengths. We compare the abundances of $2^{12}$=4096 possible 12mer sequences in positions 1 to 6 within all 72mers and compare them to each other and abundances of 12mers in positions 1 to 7 in all 84mers. Similar results were obtained for other chains longer than 36 nt. A rectangle of very high correlations (>0.9) at the center of the table in Fig. 4b means that very similar sequences get selected at all internal positions of all chains (note that only chains longer than 36nt have such internally positioned 12mers). However, the light border of the table means that a rather different subset of 12mers gets selected in the first and the last position of a multimer. Whatever the nature of selection pressure acting on these 12mers, it is consistent across oligomers of different lengths as manifested by the high correlation in the lower left and the upper right corner of the table in Fig. 4b.

A simple hypothesis comes to mind: a strand is prolonged and grows in this random sequence templated ligation system as long as the sequences attached to it share similar sequence motifs resulting in high values of sPCC for all internal 12mers. But when a 12mer sequence that is similar to the start- or end-subsequence is attached, the growth in that direction stops.

Comparison of abundances of internal 12mers in A-type and T-type subpopulations predictably yielded no positive correlation and in fact resulted in a slight negative correlation (see SI section 11). However, abundances of reverse complements of sequences from the T-type subpopulation are strongly correlated with those of the A-type resulting in a sPCC matrix similar to that shown in Fig. 4b (see SI-Fig. 12). Therefore, chains in two groups (A-type and T-type) show a considerable degree of reverse complementarity to each other. This fits the elongation and replication mechanism by templated ligation.

To further explore selection capabilities of templated ligation as a function of 12mer sequences in the initial pool we conducted three additional experiments referred to as “Replicator”, “Random” and “Network”. The “Random” experiment started with eight randomly chosen 12 nt sequences served as a control. In the “Replicator” experiment the pool consisted of eight 12 nt sequences artificially designed for efficient elongation (see below). In the “Network” experiment we populated the pool with eight naturally selected 12 nt sequences commonly found as subsequences of long strands in our original ligation experiment with 4096 12mers. To identify these 12mers, we built a network of the most common 12mers found in A-type oligomers with length of more than 48 nt. This network does not include the first and the last 12mers, in a multimer as those are known to be statistically different from the internal ones (see Fig. 4b). The circles in Fig. 5a represent unique 12 nt subsequences while their size describes their Z-scores quantifying their abundance in long chains. The width of the connecting line describes the probability that two subsequences are found one after another in a multimer. The same is done for T-type sequences (Fig. 5b). This representation of a polymer is known as de Bruijn graph[41] and has been commonly used in DNA fragment analysis and genome assembly[42] and more recently in the context of templated ligation[17].

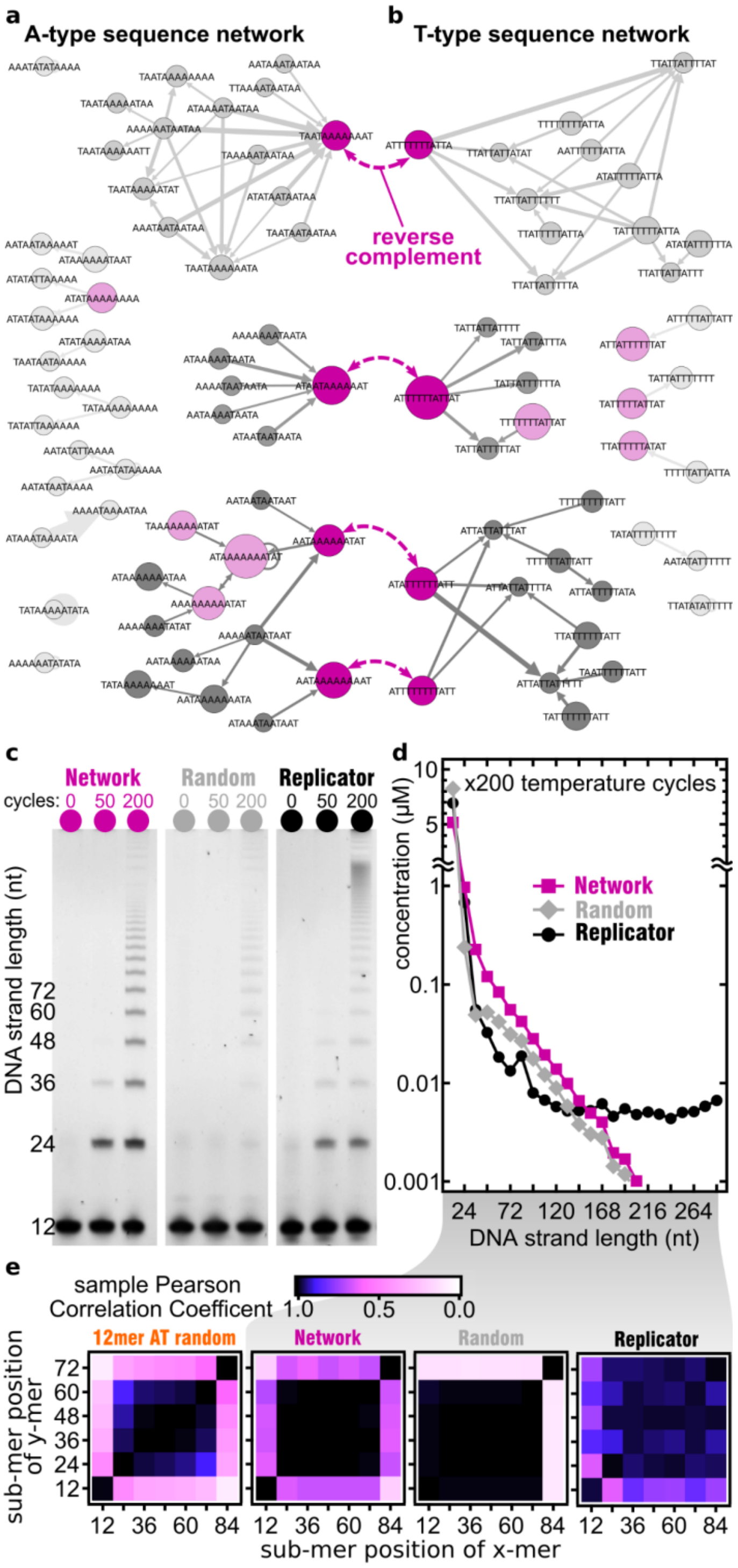

***Fig. 5, testing self-selection with custom sequence pools.***

***a*** *The de Bruijn graph of overrepresented sequence motifs between consecutive 12mers found in long oligomers. All internal junctions of A-type sequences >48 nt are shown, except the first and the last. All analyzed strands have a Z-score >30 and are sequenced at least 20 times.*

***b*** *The same de Bruijn graph but for T-type sequences with Z-score >15 and sequenced at least 10 times. Four pairs of most common reverse complementary 12mers are connected by purple dashed arrows. In each network three families with distinctly similar patterns are observed, that each include at least one of the complementary strands. Node sizes reflect relative abundance of 12mers, edge thickness denotes the Z-score of the junction between nodes it connects. Light and dark magenta-colored nodes are eight most abundant 12mers in each of two networks.*

***c*** *PAGE images of templated ligation of three different samples of 12mers after different number of temperature cycles (columns):* ***"Replicator"****: four substrate 12mers and four template 12mers artificially designed for templated ligation, as explained in SI, "****Random****": eight random sequence 12mers randomly selected from the 4096 possible AT-only 12mers, "****Network****": four most common 12mers from A-type and another four of T-type shown in dark magenta color in a).*

***d*** *After 200 temperature cycles, the "Replicator" shows a consistently higher product concentration for all lengths followed*

*by the "Network" sample and then by the "Random" subsamples. In the "Network" and "Random" samples the length distribution above 48nt is well described by an exponential distribution as predicted in Ref [39].*
***e*** *Pearson correlation matrices between 12mer abundances within 72mers and 84mers in each sample (same as in Fig. 4b). While the pattern of correlations in the "Network" sample (second from left) resembles that shown in Fig. 4b (reproduced in the leftmost subpanel), the "Random" sample (second from right) singles out the last 12mer but not the first one. The "Replicator" sample (the rightmost subpanel) has its own distinct self-similar pattern of correlations.*

De Bruijn networks in Fig. 5a break up into several clusters connecting 12mers with similar subsequences at junctions (TAA-TAA in the top cluster marked by a dark-magenta node, ATA-ATA in the middle one, and AAT-AAT in the bottom one). Note that these three common junction subsequences are all related via template shifts. The most common subgraphs found in the A-type network and mirrored among their reverse complements in the T-type network. This pattern is consistent with selection driven by templated ligation (see SI section 19). Among the eight most common subsequences in the A-type network (light and dark magenta nodes in Fig. 5a), four (dark magenta nodes) had a reverse complement among the eight most common subsequences of the T-type network (light and dark magenta nodes in Fig. 5b). These sequences were chosen as the pool of eight 12mers in the "Network" sample. The "Random" sample consisted of eight 12mers which were randomly chosen from the 4096 possible AT-only 12mers. The "Replicator" sample consisted of eight strands that were built to form three-strand complexes that resemble the assumed first ligation reaction in the pool (SI section 17.1).

The length distribution of oligomers (Fig. 5d) with concentrations quantified from the PAGE gel image (Fig. 5c) shows that the "Network" sample produced the most product, as the remaining 12mer sequence concentration was reduced below two other samples down to almost 5 µM. The length distribution in both "Random" and "Network" samples is well described by a piecewise-linear distribution predicted in Ref [39]. For short product lengths ranging between 48mers up to 136mers the "Random" sample produced more oligomers than the "Replicator" sample. However, for even longer strands, the "Replicator" sample generated the largest number of really long strands since its length distribution reached a plateau around 120mers. This is probably due to the nature of the eight-sequences pools used here with the "Replicator" one made to form well aligned dsDNA that can be properly ligated. According to NUPACK[43], 12mers in the "Random" sample should not form any complexes that could be subsequently ligated by the TAQ ligase. However, our results shown in Fig. 5c prove the existence of extensive ligation even in the "Random" sample. Presumably, it was initially triggered by small concentration of complexes formed with low probability, which were subsequently amplified due to the exponential growth of longer strands in our experiment, just like in the "Network" sample.

## DISCUSSION

We experimentally studied templated ligation in a pool of 12mers made of A and T bases with all possible sequences ($2^{12}$=4096), subjected to multiple temperature cycles. To accelerate hypothetical spontaneous ligation reactions operating in the prebiotic world, we employed TAQ DNA ligase in our experiments. This process produced a complex and heterogeneous ensemble of oligomer products. By performing the "next generation sequencing" of these oligomers, we found that long strands in this ensemble have a significantly lower information entropy compared to a random set of oligomers of the same length. This effect became increasingly more pronounced for longer oligomers (Fig. 2e). The overall reduction in entropy was in line with the theoretical prediction obtained within a simplified model of template-based ligation[17]. In that model, the reduction of entropy was due to "mass extinction" in sequence space, with only a very limited (though still exponentially large) set of survivor sequences emerging. In the present experiment related variation in abundances of different sequences did develop but didn't proceed all the way to extinction.

Several patterns can be easily spotted in the pool of surviving sequences. In particular, multimer strands predominantly fell in one of two groups: A-type or T-type each characterized by about 70 % of either base A or T (Fig. 2c, d). The initially single-peaked approximately binomial A:T-ratio distribution in random monomers changed into a bimodal one in longer chains. We attribute this separation into two subpopulations to the fact that such composition bias suppresses the formation of internal hairpins and other secondary structures. The self-hybridization reduces the activity of both template and substrate chains leading to a lower rate of ligation. The adaptation by separation into two subpopulations was reproduced by a kinetic model in which activities of reacting strands were corrected for hairpin formation, with realistic account for its thermodynamic cost. This model produced a bimodal distribution of A-content in 24mers, in qualitative agreement with the experimental data. Furthermore, the eventual distribution of longer oligomer lengths could be successfully captured by the maximum entropy distribution, subject to the constraint of fixed average composition of A- and T-type subpopulations. Another remarkable observation is that although formation of hairpins was suppressed through the mechanism above, a small but noticeable fraction of oligomers have extremely long stretches of internal hairpins. The likely mechanisms of their formation are either ligation of a pair of nearly complementary chains from A-type and T-type subpopulations, or self-elongation of such oligomers.

Another common pattern was a distinct AT-alternating pattern around the ligation site, as can be seen in Fig. 1b. Those AT-alternating motifs first appeared in 24-mers, and remained very common in longer chains. These features accounted for some of the reduction in sequence entropy, but did not account for all of the selection at ligation sites, where, as demonstrated by the Z-score analysis, a rich ligation landscape has developed (Fig. 4a, b). Not only some 12mers within longer chains were far more abundant than average, but there were also pairs of those that preferentially follow each other, as demonstrated by de Bruijn graphs in Fig. 5a, b.

We selected a subset of eight pairs of mutually complementary 12mers that appeared anomalously often within longer chains and were well connected within the de Bruijn graph. Using this “Network" subset as a new starting pool, we repeated the temperature-cycling experiment, and compared it to two other reference systems. One of them were eight randomly selected 12mers, the other was artificially designed to promote self-elongation. The resulting multimer population in two out of three of these pools followed a near perfect exponential length profile (Fig. 5d). The random pool resulted in a similar behavior to the network one but with significantly lower overall concentration of long chains. Both results are in an excellent agreement with theoretical predictions of reference[39]. A higher concentration of long chains generated by network 12mers indicates better overall fitness of this set compared to random 12mers. The “Replicator” set did produce a large number of very long products, presumably by a different mechanism, but a significantly smaller number of products with short and medium lengths. This indicates lower autocatalytic ability in both “Replicator” and “Random” sequence pools when compared to the “Network” pool.

For emergence of life on early earth, random oligomers needed to act in an evolution-like behavior. Here, we followed templated ligation of random 12mer strands made from two bases under temperature oscillations. Despite its minimalism, the system contains all elements necessary for Darwinian evolution: out of equilibrium conditions, transmission of sequence information from template to substrate strains, reliable reproduction of a subset of oligomer products and selection of fast growing sequences in the process. At the dawn of life, pre-Darwinian dynamics would have been important to push prebiotic systems towards lower entropy states. Such pre-selection for catalytic function could have paved the way towards eventual emergence of life.

# METHODS

## Nomenclature

***Oligomer***: a product from the templated ligation reaction with a length of a multiple of 12 nt. ***Subsequence***: 12mer long sequence in between two ligation sites or in the beginning or end of a multimer. ***Submotif***: a sequence of a certain length $x$. In contrast to a subsequence, a submotif can start at any position in a mono- or oligomers, not only at ligation sites, or the sequence start. ***Ligation site:*** in particular, the bond between two monomer or multimer strands. In context of sequence motifs, it refers to the region around this bond (±1 to 6 bases).

## Ligation by DNA ligase

For enzymatic ligation of ssDNA a TAQ DNA ligase from *New England Biolabs* was used. Chemical reaction conditions were as stated by the manufacturer: 10 µM total DNA concentration in 1x ligase buffer. The ligase has a temperature dependent activity and is not active at low (4-10 °C) and very high temperatures (85-95 °C). In our experimental system DNA hybridization characteristics are strongly temperature dependent, as shown in the SI. We expect this to have stronger influence on the overall length distribution and product concentrations than ligase activity, as the timescale of hybridization is significantly longer than the timescale of ligation (compared in SI). The manufacturer provides activity of the ligase in units/ml, specifically: “one unit is defined as the amount of enzyme required to give 50 % ligation of the 12-base pair cohesive ends of 1 µg of BstEII-digested λ DNA in a total reaction volume of 50 µl in 15 minutes at 45 °C”.

## Design of the random sequence pool

The use of a DNA ligase enables very fast ligation with low error rate. But not every DNA system is suitable for templated ligation. As stated by the manufacturer, the TAQ ligase does not ligate overhangs which are 4 nt or shorter. Therefore, the shortest possible length of strands is 10mer, opening up $4^{10} > 10^{6}$ different monomer sequences. The resulting pool cannot be sequenced to a reasonable extend. We artificially reduced the sequence space by limiting sequences to only include bases adenosine (A) and thymine (T). 10mer strands with random AT sequence have too low melting temperature, in a range where the ligase is not active (compare SI). We found 12mers with random AT sequences to successfully ligate and to produce longer product strands due to their elevated melting temperature. The monomer sequence space is $2^{12}$=4096 is not too large, so that we were able to completely sequence it multiple times.
The DNA was produced as 5’-WWWWWWWWWWWW-3’ with a 5’ POH modification by *biomers.net*. “W” denotes base A or T with the same probability. We analyze the “randomness” of this pool in the SI.

## Temperature Cycling

Temperature cyclers *Bio-Rad* T100, *Bio-Rad* CFX96, *Analytik Jena* qTOWER$^3$ and *Thermo Fisher Scientific* ProFlex PCR System were used to apply alternating dissociation and ligation temperatures to our samples. The dissociation temperature of 75 °C was chosen, to melt short initially emerging ssDNA of up to 36mer. In the SI we also show how a variation of the dissociation temperature changes multimer product distribution in a random sequence templated ligation experiment. Lower dissociation temperatures enable us to run several thousand temperature cycles, as the stability of the TAQ DNA ligase is reduced substantially for longer times at 95 °C. Time resolution experiments with PAGE-analysis demonstrated ligase activity even after 2000 temperature cycles for a dissociation temperature of 75 °C. In experiments screening the ligation temperature (see SI), we found that for ligation temperatures of 25 °C the product length distribution is exponentially falling. For higher ligation temperatures such as 33 °C we find more long sequences, but almost no 24mer and 36mer sequences. For sequenced samples we chose a ligation temperature of 25 °C because the library preparation kit is better suited for shorter DNA strands. In sequencing data for samples with 33 °C the yield was very low, but the results are similar to the sequencing data of samples with 25 °C ligation temperature, but with comparably worse statistics. For dsDNA dissociation in each temperature cycle the corresponding temperature is held for 20 s with subsequent 120 s at the ligation temperature.

## Sequencing by Next Generation Sequencing (NGS)

For sequencing we used the Accel-NGS 1S Plus DNA Library Kit from *Swift Biosciences*. The sequencing was done using a HiSeq 2500 DNA sequencer from *Illumina*. The kit was used as stated in the manufacturer's manual. All volumes were divided by four to achieve more output from a limited supply of chemicals. Library preparation was done in four steps: first a random sequence CT-tail was added to the 3' end of the DNA by (probably, the manufacturer does not give information about this step) a terminal transferase. In a single 15 min ligation step the back primer sequence (starting with AGAT…) was ligated to the 3' end of the random CT-stretch. In the second step a single cycle PCR was used to produce the reverse complement and to leave double stranded DNA with a single A overhang. Step three ligated the start primer to the 5' end of the DNA. Step four added barcode indices to both ends of the DNA by a PCR reaction. This step was done several times to result in the desired amount of DNA for sequencing.

## Sequence Analysis

Demultiplexing was done by a standard demultiplexing algorithm on servers of the Gen Center Munich running an instance of Galaxy[44] connected to the sequencing machine. *Illumina*-sequencing creates three FASTA-files, listing the front and the back barcodes and the read sequence, for each lane of the flow cell. The demultiplexing-algorithm matches the barcodes of the prepared library DNA to the read sequence and produces a single FASTA file including the read quality scores.

The sequence-data was analyzed with a custom written *LabVIEW* software. The main challenge was to separate the read sequences from the attached primers. The start primer is automatically cut in the demultiplexing step. The end primer is cut with an algorithm based on regular expression (RegEx) pattern matching. With RegEx we first search for multiples of the monomer length. If these structures were followed by at least four bases of C or T followed by the sequence AGAT we concluded that we found a relevant sequence. The 3'-primer was cut and the resulting sequence saved for analysis.

RegEx for searching AT random sequences:

```
(^[ATCG]{12}|[ATCG]{24}|[ATCG]{36}|[ATCG]{48}|[ATCG]{60}|[ATCG]{72}|[ATCG]{84})(?=([CT]{4,}AGAT))
```

RegEx for selecting a maximum of X false reads of G or C in random sequence AT samples: `^(?!(?:.*?(G|C)){X,})^([ATCG]{12,})`. The sequenced library may have primer-primer dimers and oligomers as well as partial primers that were falsely built in the library preparation step. As the SWIFT kit is made for longer sequences by design, shorter sequences such as 12mer in our study may have lower yields and larger error rates for the library kit chemistry. Therefore, the inclusion of sequences with a single or multiple false reads can improve the statistics, as long as submotifs with obviously faulty reads are ignored in the analysis.

# BACKMATTER

**Competing Interests**

The authors declare that they have no competing interests.

**Author's contribution**

P.W.K. performed the experiments, prepared the library for sequencing, performed the demultiplexing, the analysis, programmed the analysis software, analyzed the data, drafted and wrote the manuscript. A.V.T and S.M. performed the theoretical analysis and analyzed the data in context of their already published theoretical work, drafted graphs, drafted and wrote the manuscript. D.B. contrived the experiment, guided the experimental progress, analyzed data and drafted the manuscript.

**Acknowledgements**

The authors would like to acknowledge funding by the Deutsche Forschungsgemeinschaft (DFG, German Research Foundation)– Project-ID 201269156 – SFB 1032, the Advanced Grant (EvoTrap #787356) PE3, ERC-2017-ADG from the European Research Council, CRC 235 Emergence of Life (Project-ID 364653263) and the Center for NanoScience (CeNS). We would like to thank Ulrich Gerland, Tobias Göppel, Joachim Rosenberger and Bernhard Altaner for their helpful remarks and discussions about hybridization energies, baseline corrections and interpretation of multimer product distributions. P.W.K and D.B. thank Stefan Krebs and Marlis Fischalek at the Gene Center Munich for help with the library preparation and the sequencing the samples and Annalena Salditt and Filiz Civril for comments on the manuscript. This research was partially done at, and used resources of the Center for Functional Nanomaterials, which is a U.S.
DOE Office of Science Facility, at Brookhaven National Laboratory under Contract No.~DE-SC0012704.

# Ligation of random oligomers leads to emergence of autocatalytic sequence network

Patrick W. Kudella[1], Alexei V. Tkachenko[2], Sergei Maslov[3,4], Dieter Braun*[1]

[1] Systems Biophysics and Center for NanoScience, Ludwigs-Maximilian-Universität München, 80799 Munich, Germany
[2] Center for Functional Nanomaterials, Brookhaven National Laboratory, Upton, New York 11973, USA
[3] Department of Bioengineering, University of Illinois at Urbana-Champaign, 1270 Digital Computer Laboratory, MC-278, Urbana, Illinois 61801, USA
[4] Carl R. Woese Institute for Genomic Biology, University of Illinois, Urbana-Champaign, Illinois 61801, USA

# Supplementary Information

# 1. Table of Contents

## 2. Polyacrylamide gel electrophoresis

For analyzing the dynamics and product yield of the random sequence ligation we use polyacrylamide gel electrophoresis (PAGE) with SYBR gold post-staining. The gels are 15 % acrylamide and are run in a solution of 50 % urea and 1x TBE buffer at about 50 °C posing denaturing conditions.

The gel is mixed from the *Roth* Rotiphorese DNA sequencing system. One 0.75 mm thick gel with a 15 tooth comb needs about 5 ml gel mixture which contains 3 ml gel concentrate, 1.5 ml gel diluent, 0.5 ml buffer concentrate, 25 µl APS and 2.5 µl TEMED. After 30 min of pre-run at 400 V, the gel pockets are loaded with a total of 4 µl of sample made from 0.89 µl of 10 µM sample and 3.11 µl of 2x loading dye (for about 10 ml add 9.5 ml formamide, 0.5 ml glycerol, 1 µl EDTA (0.5 M), and 100 µl Orange G dye (e.g. from *New England Biolabs*)). The sample is drawn into the gel in a first step of the run with 50 V for 5 min, then the gel electrophoresis is run for about 30 min at 300 V.

After the run, gels are submersed into 50 ml of 1x TBE buffer with 5 µl of 10.000x *SYBR Gold Nucleic Acid Gel Stain* from *Thermo Scientific* for 5 min. The stained gel is washed in 1x TBE buffer two times and imaged in a *bio-rad ChemiDoc MP* System.

Analysis of the gel images are done in self-written *LabVIEW* code, annotations are made in *GIMP* and *inkscape*.

## 3. Concentration Measurement in PAGE Gels by Image Analysis

We developed a LabView program for detailed analysis of PAGE images. The main problem with quantification in PAGE is the inhomogeneous fluorescence for ssDNA and dsDNA, as well as base-order. In our experiments all possible products are made from the same length monomers that are only have A and T as. The necessary prerequisite for the tool is a baseline: this lane is loaded with the monomers and buffer only. This baseline sample did not experience temperature cycling but was kept at constant 4 °C in the fridge. All other samples are compared to this sample.

In the first step, gel images are loaded in the tool and the outermost lanes are marked with cursors. The program automatically selects the lanes and space in between the lanes for background correction. For every lane the intensity over the gel position is calculated and corrected with the average intensity taken from the inter-lane areas left and right of the sample lane.

The intensity data is then normalized: the total intensity in every lane must be the same as the total concentration of the monomer pool in the baseline lane. The peak areas are marked with cursers and integrated with the simple trapezoidal method. For the final concentration estimation, the total intensity of the $x$th peak is divided by $x$ – as a $x$-mers' intensity is $x$ times greater than that of a monomer.

Performing the same experiment several times and calculating the concentration from the different gels each time enables us to calculate a standard deviation, as shown in SI-Fig. 1. The deviation of the six samples is small and the traces for the different ligation temperatures can be easily distinguished.

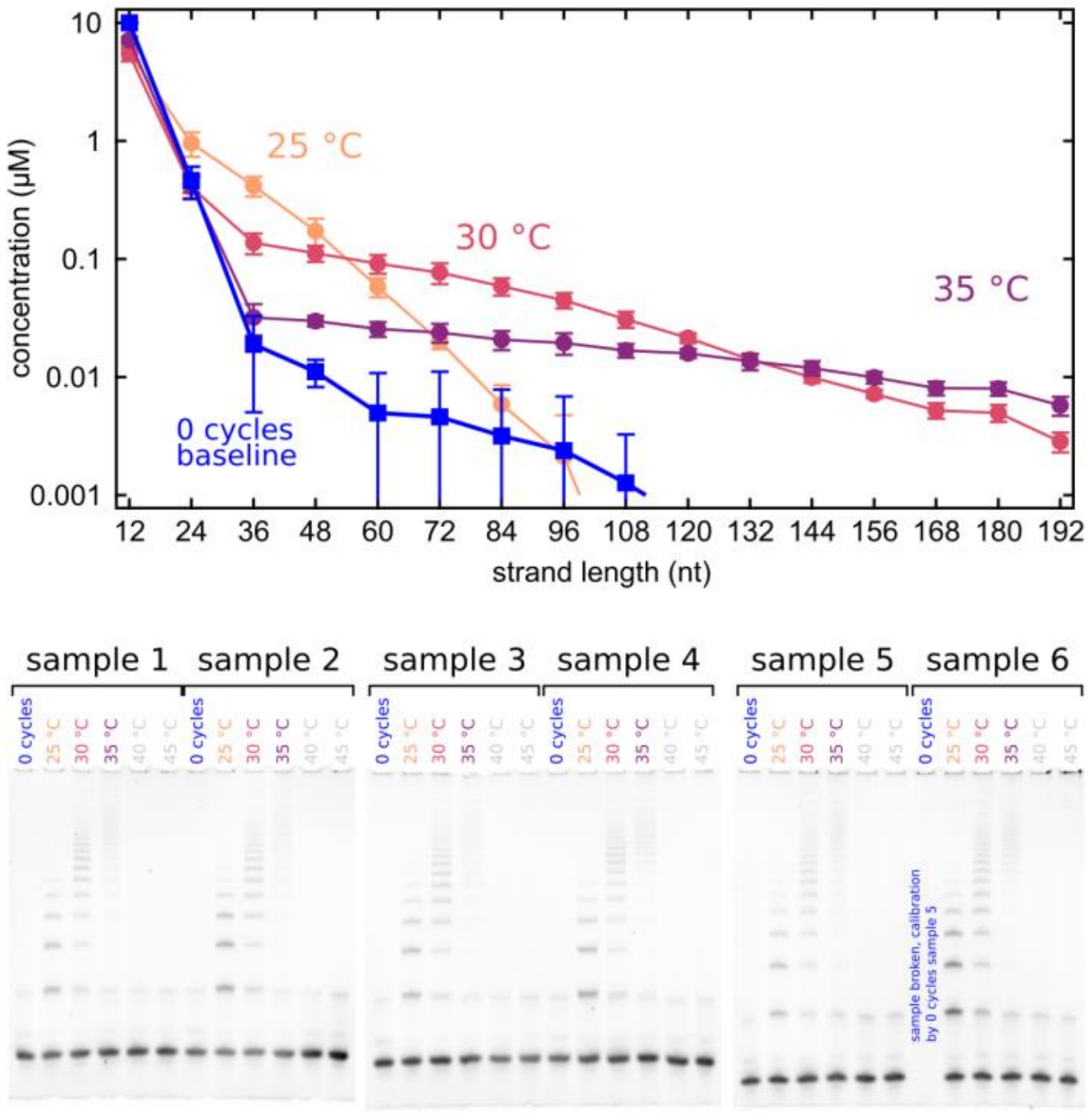

*SI-Fig. 1, **reproducibility experiment for concentration estimation tool,***
***top:** Calculating the concentration from each experiment enables the calculation of an average concentration estimation and a standard deviation error.*
***bottom:** Three gels with a total of six experiments show very similar structure.*

## 3.1. Baseline of Gels

All gels show two different distinct artefacts. First, there is a small peak directly after the 12mer peak. This peak is only visible if the sample contains the ligase buffer. We suspect the SYBR gold post-staining also dyes a component of the buffer. In SI-Fig. 2 the gel and intensity analysis both show this peak. In comparison a sample with just DNA and MilliQ water does not show this peak.

The second peak unfortunately runs at a length where one would expect the 24mer products of the ligation reaction. This peak is always visible at the same location with the same intensity. This is probably an artefact from DNA synthesis. As the DNA is synthesized starting from the 3'end, it can occur, that a second DNA backbone is synthesized to the first base. Those structures would run comparable to a 24mer, but probably don't take part in the ligation reaction. It can, however, not be analyzed by NGS, as the library preparation chemistry is sensitive to DNA backbone errors.

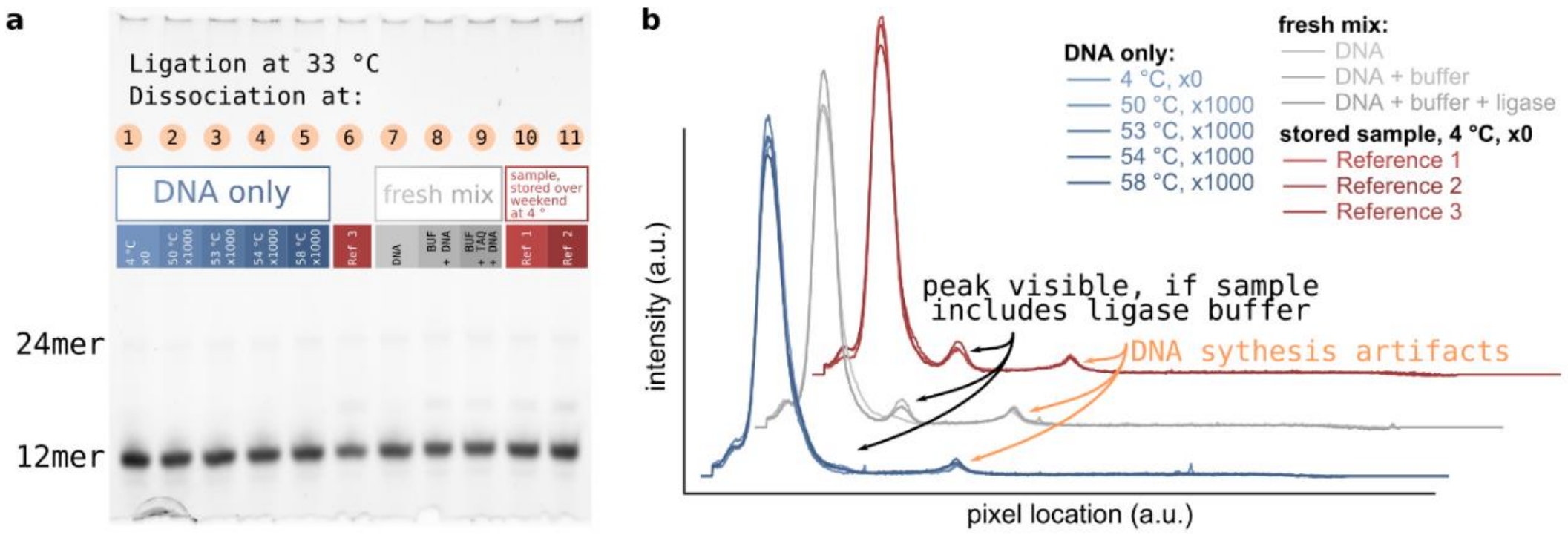

*SI-Fig. 2, **PAGE of AT-only random sequence pool with and without ligase for baseline comparison:***
***a** PAGE of AT-only random 12mer DNA with different buffer conditions. 1) DNA only in MilliQ water at 4 °C in the fridge for ~60 h. 2) DNA only in MilliQ water, 1000 temperature cycles between 33 °C and 50 °C. 3) DNA only in MilliQ water, 1000 temperature cycles between 33 °C and 53 °C. 4) DNA only in MilliQ water, 1000 temperature cycles between 33 °C and 54 °C. 5) DNA only in MilliQ water, 1000 temperature cycles between 33 °C and 58 °C. 7) fresh solution of DNA in MilliQ water, immediately mixed with loading dye and quenched. 8) fresh solution of DNA in MilliQ water and ligase buffer, immediately mixed with loading dye and quenched. 9) fresh solution of DNA in MilliQ water, ligase buffer and ligase, immediately mixed with loading dye and quenched. 10), 11) and 6) AT-only random DNA in ligase buffer and ligase, stored at 4 °C in the fridge for ~60 h.*
***b** Baseline-corrected intensity plots of the gels: all lanes show the same artifact at a position where 24mer DNA would run. This might be an artifacts of the synthesis at the 3' end of DNA. The small peak close to the 12mer-peak is only visible if the ligase-buffer is in the sample. All of those artifact-peaks are very similar across different experimental conditions.*

# 4. NUPACK simulation comparing 12mer and 10mer AT complexes

As mentioned in the main manuscript double stranded complexes in a conformation that can be ligated by the TAQ DNA ligase are needed in order for the experiment to work. Taking later results as the input for the most probable sequences for a complex of three strands we use NUPACK to calculate an approximation to the melting curve. SI-Fig. 3a shows the three sequences with lengths 10 nt and 12 nt. SI-Fig. 3b shows the fraction of unpaired bases as a function of temperature. The concentration of strands is set to 0.0098 µM and 0.0024 µM which is equivalent to their concentration as if they had the same abundance as in a 10mer (respectively) 12mer AT-random sequence pool.

In the temperature range where ligation reactions are possible because the ligase is active, there is a significantly lower number of paired bases marking a lower number of double stranded DNA for the 10mers. With an addition of only 2 bases per stand, the total sequence space only grows by a factor of 4. And NUPACK suggest a higher number of double strands at temperatures of 25 °C to 33 °C – conditions similar to the experimental ligation conditions in the main manuscript.

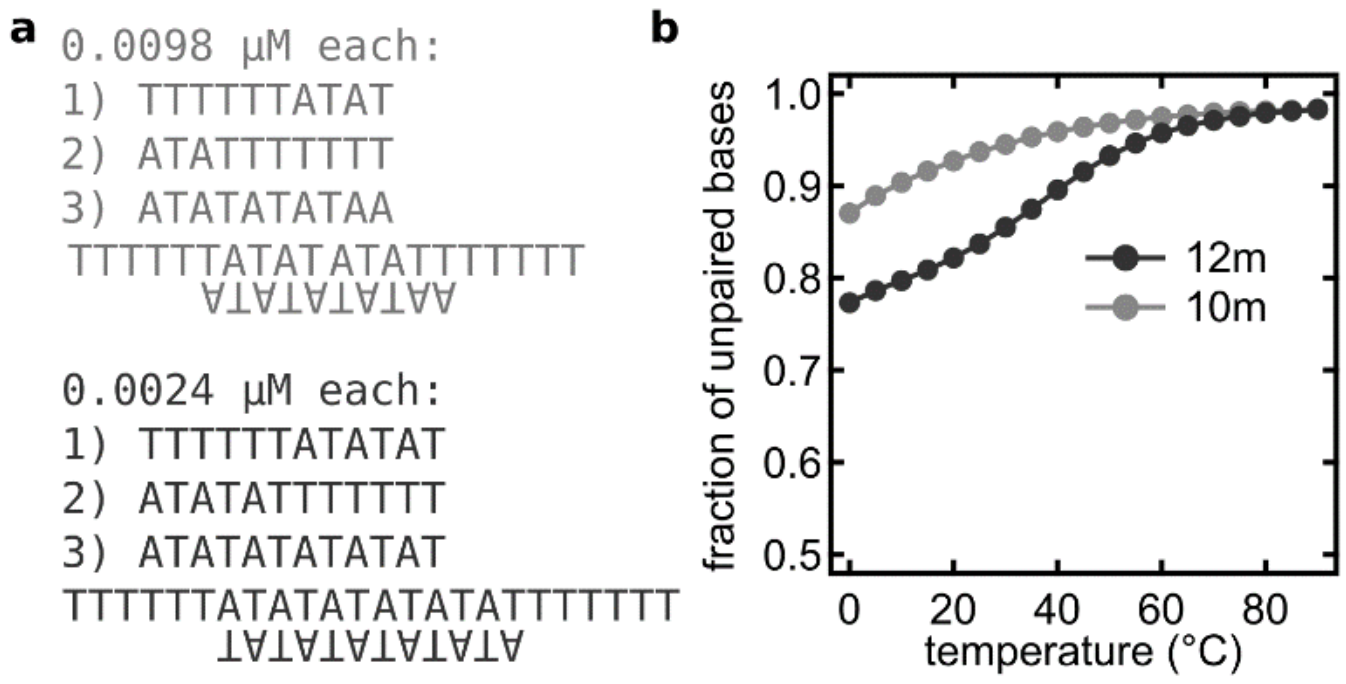

*SI-Fig. 3, **NUPACK simulation for 12mer and 10mer AT sequences duplex formation:***
***a** 10mer and 12mer long strands that form triplet structures suitable for TAQ DNA ligation.*
***b** The NUPACK tool gives a fraction of unpaired bases over temperature for both systems. This is comparable to a melting curve: the lowest possible value on the y-axis is about 0.33 because in ideal conformation there are 24 of the 36 bases paired, leaving 1/3 unpaired. At the ligation temperature of 33 °C there is a significantly larger amount of paired 12mers than 10mers.*

# 5. Random sequence pool parameter space

The pool made from 12mer AT-random sequence DNA strands produces oligomers of different lengths by the templated ligation reaction under temperature cycling. The dynamics of the system, best visualized by PAGE, depends on several parameters discussed below.

## 5.1. Random sequence pool ligation dynamics as function of temperature

The dynamics of formation of longer strands from the monomers in the pool is highly dependent on the temperature of ligation and dissociation. As a rule of thumb, higher temperatures of ligation produce more long strands with non-exponentially decreasing character. Higher dissociation temperature reduces the amount of long sequences dramatically. In SI-Fig. 4 four gels show the described behavior. SI-Fig. 4a shows 12mer AT-only random sequence pool products for varying ligation temperatures and a dissociation of 75 °C. The sample at 25 °C shows a strong exponential decay of longer sequences. For 30 °C the exponential decay is weaker and the amount of long sequences is higher. For 35 °C only long sequences emerge and short 24mer and 36mer sequences are absent in the gel. For high temperatures of 40 °C and above no product emerges. In SI-Fig. 4b the same sample is run with a dissociation temperature of 95 °C. The overall behavior is similar, but the amount of long strands severely reduced.

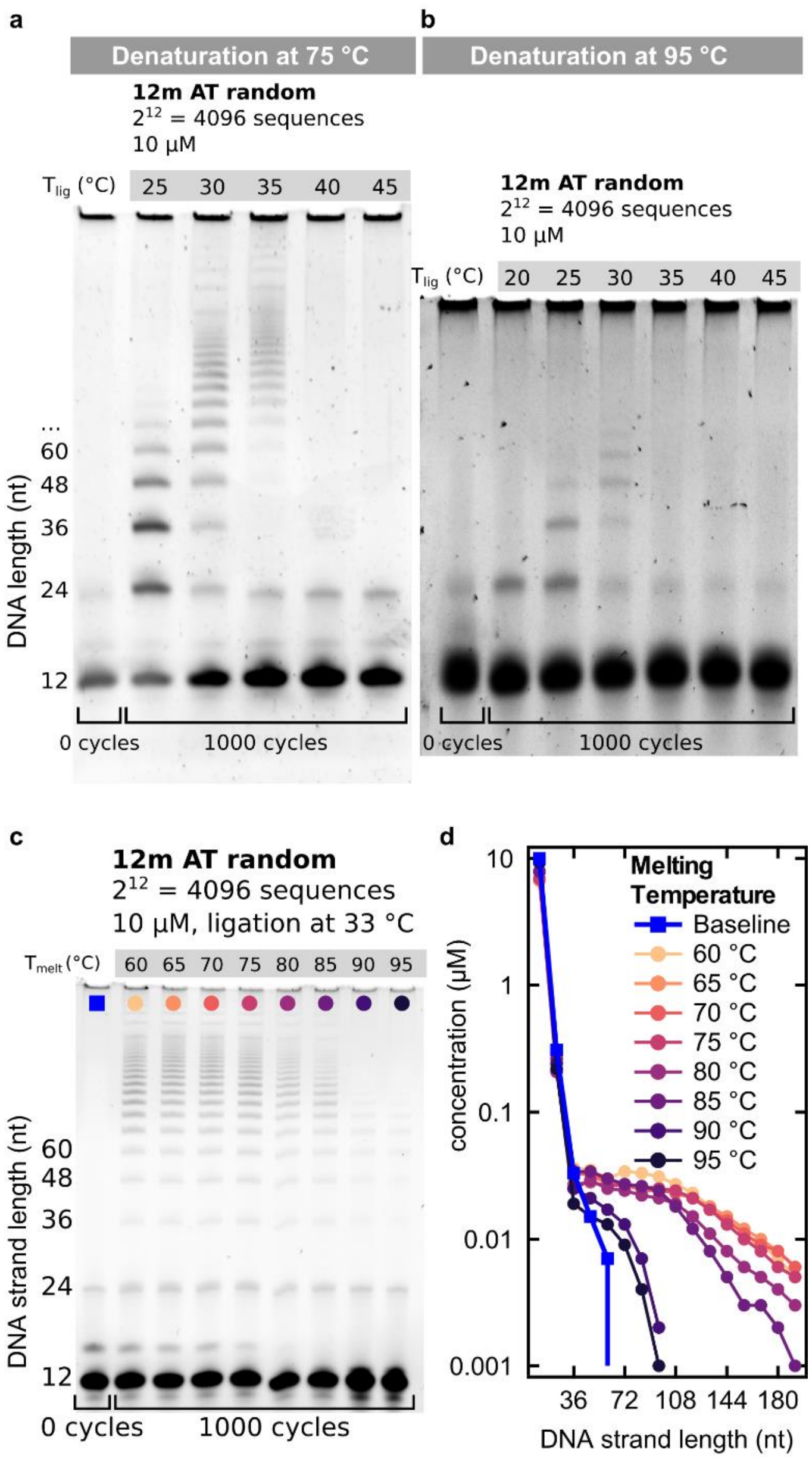

*SI-Fig. 4, **PAGE of 12mer AT only random sequence pool, cycled for 1000 times, different temperature cycle conditions:***
***a** The dissociation temperature is 75 °C the left-most line is the reference sample with zero temperature cycles. The temperature given above each lane is the temperature of the ligation reaction.*
***b** The dissociation temperature is 95 °C the left-most line is the reference sample with zero temperature cycles. The temperature given above each lane is the temperature of the ligation reaction.*
***c & d** With a fixed ligation temperature of 33 °C the dissociation temperature is varied in between lanes from 60 °C to 95 °C. The shoulder of short and medium sized oligomers is retained up until 85 °C but the concentration of long oligomers is lower, the higher the dissociation temperature.*

## 5.2. Full random ATGC 12m sequence pool

Despite the comparably easy analysis of binary sequence data in AT-only or GC-only experiments, the emergence of potential motives from full random, four bases DNA might be interesting. In preliminary results under similar experimental conditions as described above, we could not detect oligomer products emerging from 12mer ATGC-random pools. Comparing the sequence space reveals the vast difference: $4^{12}≈16.8*10^{6}$, for four bases and $2^{12}=4096$ for two bases. Assuming a linear correlation (lower boundary) of temperature cycles to sequence space for the emergence of a well observed oligomer product distribution would mean an increase of experimental time from about 60 hours for 1000 temperature cycles to 204.800 hours (>23 years). With changes to the experiment it might be possible to decrease the experimental time further, but the system will have either reduced complexity, or less freedom to form complexes.

Anyhow, the underlying results found in this work might very well still be valid: suppression of self-folding, emergence of specific sequences at ligation sites due to better binding geometry or higher binding energy, suppression of poly-G sequences due to their stickiness might all be found in a potential full random templated ligation samples.

### 5.3. Comparison with Toyabe/ Braun

The 2019 study[1] of Toyabe and Braun explored possible cooperation of selected DNA motives of length 20 nt. The strands were designed to build networks, that could withstand a simulated decay (serial dilution) when fed with the motive-monomers. The ligation was also done with the TAQ DNA ligase from *new england biolabs*. For the dissociation temperature they chose 95 °C, the highest possible temperature where the ligase can survive for short time and DNA does not denaturate, while double strands are dissociated. They kept the dissociation temperature time as short as the PCR temperature cycler allows (1 s). The ligation temperature was selected as 67 °C, the melting temperature of the motive-monomers. They used three different submotives (a, b, c) and their respective reverse complements ($\bar{a}$, $\bar{b}$, $\bar{c}$), and different oligomers to start the reaction, like $ab + \overline{ba}$ noted as AB together. In a typical reaction six monomers plus 2 for each oligomer (two different 40 nt dimers and one 60 nt trimer) were spanning a sequence space of twelve different sequences.

Although the total sequence space for 12mer “monomers” in the experiment in the main manuscript is easy to calculate ($2^{12}$=4096) and as shown below also a good estimate for the actually present DNA strands in the random sequence mix, the comparison is not straight forward. In both experiments, the hybridization is the actually interesting mechanism. In Toyabe and Brauns work, the strands are designed to only hybridize to their reverse complement and without any overhangs. In contrast, strands from the AT-only random sequence pool can hybridize in multiple different configurations and a majority of those will likely inhibit ligation. The sequence space of complex formation is therefore significantly larger than the increase in sequence species might suggest. This slows the overall reaction rate substantially and explains, why the random sequence templated ligation reaction needs longer cycle times.

The melting temperature of a dsDNA strand depends on several parameters like the amount of paired bases, the GC-content and stacking interactions. In Toyabe and Brauns work all “monomers” were designed to have the same melting temperature. In the random sequence pool, dsDNA complexes presumably have very different dissociation temperatures, mainly due to the different amount of paired bases, dangeling ends and the actual sequence of bases.

# 6. GC only random sequence pools of different starting lengths

The main reason for using AT-only samples is the lower melting point for dsDNA. Using a GC-only random sequence pool, complexes have a higher melting temperature and the experiments then also need a higher dissociation temperature in which the ligase degrades faster. Nevertheless, the basic templated ligation elongation experiment is possible for GC-only pools. With the higher melting temperature of the 12mers, we also tried lower lengths for the random sequence pool. As explained in the methods section of the main manuscript, the dissociation temperature is a limiting factor for the shortest possible monomer length of AT-only strands. SI-Fig. 5 shows that the 8mer GC-only random sequence pool does not produce oligomers in 1000 temperature cycles. This is not surprising, as the manufacturer states 4mer overhangs are not ligated by the TAQ DNA ligase. From all other pools of lengths 9mer (remarkable, as a complex made from three 9mers must have a 5mer and a 4mer overhang – which is apparently ligated), 10mer and 12mer oligomers emerge, although way less long oligomers and less oligomers over all. The bands, especially for the monomer-length, tend to smudge into/ close to the next bands. There is also a more pronounced smear of each product band.

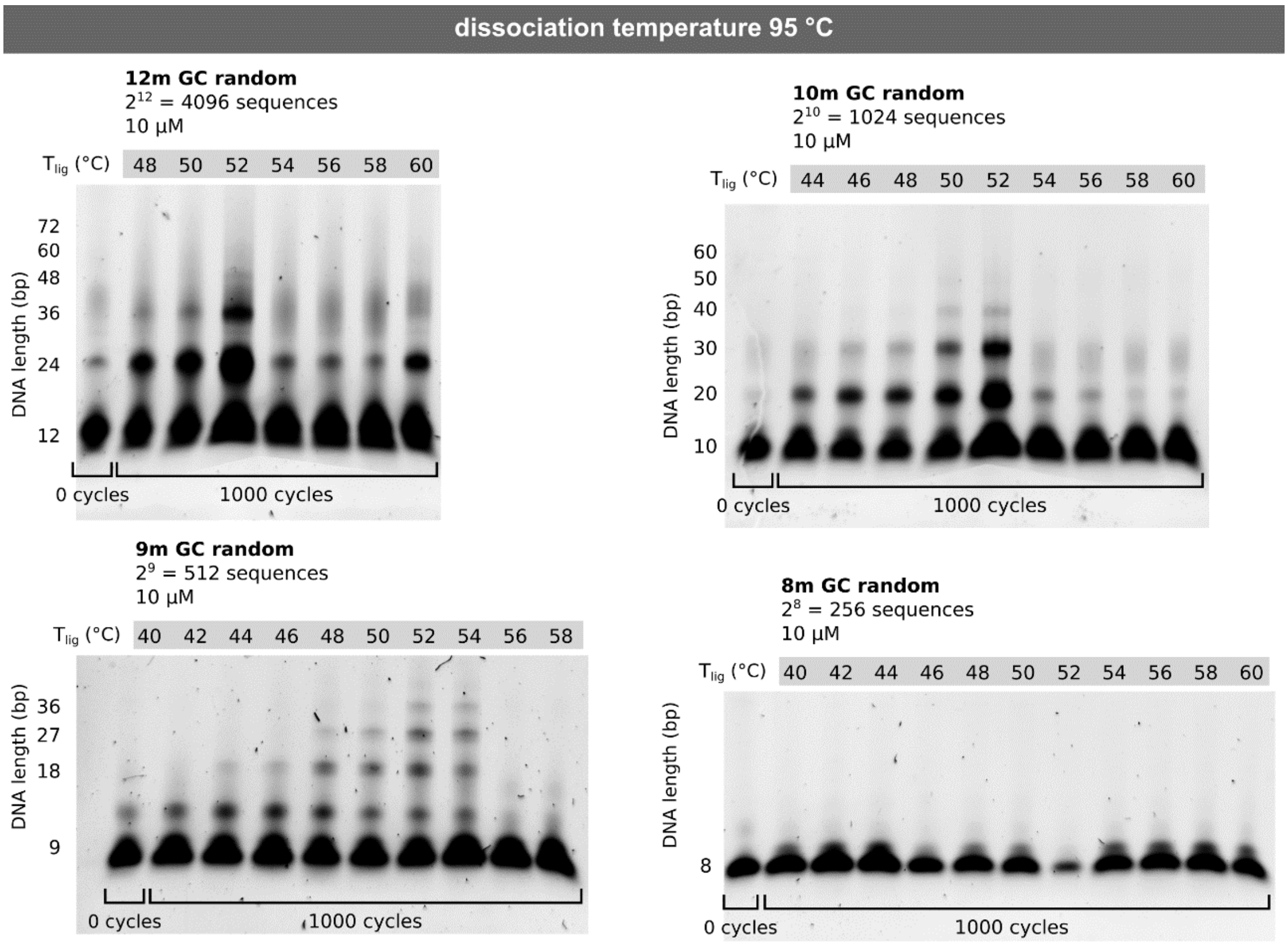

*SI-Fig. 5,* ***PAGE of several lengths of GC only random sequence pools, cycled for 1000 temperature cycles with different conditions:***
***top left:*** *12mer sample with sequence space 4096. Although using a denaturizing PAGE conditions, bands tend to smear out for longer strands.*
***top right:*** *10mer sample with sequence space 1024.*
***top right:*** *9mer sample with sequence space 512. For this length, the length distribution for reaction products looks the most like the AT only random sequence 12mer shown in the main manuscript.*
***top right:*** *8mer sample with sequence space 256. Here, the strands are too short and the ligase doesn't ligate the three-part complexes anymore.*

The best working conditions for the GC-random pools seems to be around 52-54 °C, which is significantly higher than for the AT-only random sequence pools (33 °C).

# 7. Ligation time variation

The sequence space of a pool of DNA dictates the ligation time necessary for the emergence of products. For the 12mer AT-only random sequence pool temperature cycling is necessary, as seen in SI-Fig. 6. For this sample, the increase of the ligation time in each respective temperature cycle yields more product in total. At 10 s and 20 s ligation there is no product at all. The ligation time of 120 s gives a lot of product but might not mark the maximum possible product yield. The last lane of the PAGE in SI-Fig. 6 shows that the same sample at a constant temperature of 33 °C for approximately 60 h does not significantly differ from the sample with a ligation time of 10 s.

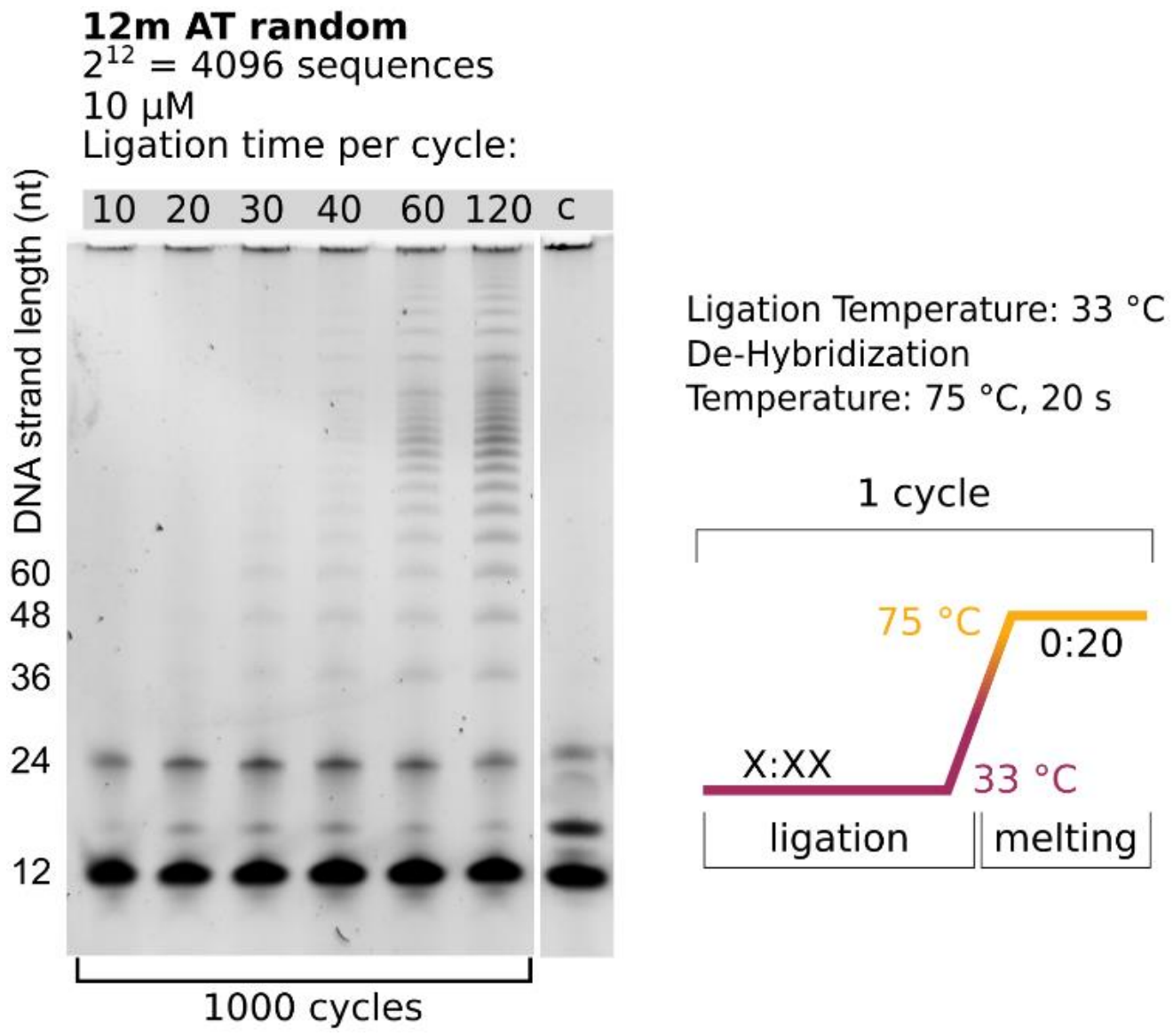

*SI-Fig. 6,* ***PAGE of AT-only random sequence pool (orange) and NN (blue) sample:***
*For longer ligation times the gel shows more product for the AT-only random sequence pool sample. The last columns indicated "c" on top mark the sample without temperature cycling. For ligation time steps of 10 s and 20 s the gels doesn't show any product at 24mer length or longer. For longer times, oligomers start to emerge. The sketch shows the temperature-cycle scheme used here.*

We chose our temperature cycling conditions optimized for the emergence of oligomer products with 20 s dissociation temperature and 120 s ligation temperature. With the heating and cooling in between those steps the entire experiment lasts about 60 h for the 1000 temperature cycles.

Additionally, as seen in section 5.1, our experimental system is strongly temperature dependent. We expect this to have a stronger influence on the overall length and product concentration distribution than the activity if the ligase.

As oligomers emerge in all but the very short ligation time steps experiments, we estimate that the timescale of hybridization due to the vast sequence space is substantially larger than the ligation time scale.

# 8. LabVIEW program for sequence analysis

Sequence analysis is predominantly performed with self-written LabVIEW programs. LabVIEW is a graphical programming language suited for fast programming of high level data structures. A LabVIEW program is called a "Virtual Instrument" (VI) and consists of the front panel including graphs, controls and tables and the block diagram including the logic of the program. Overall, LabVIEW is very similar in performance compared to other high level scripting languages but has several advantages like the by default included user interface (UI) or the possibility to store large datasets in the VI.

When an *illumina* sequencing run is completed the data is stored on a server of the Gen Center Munich (https://www.genzentrum.uni-muenchen.de/index.html) running an instance of galaxy[2]. galaxy-demultiplexing scripts are used for the demultiplexing step of the data, as described in the methods section. The demultiplexed FASTA file is then downloaded and further analysis is done with the custom-VIs.

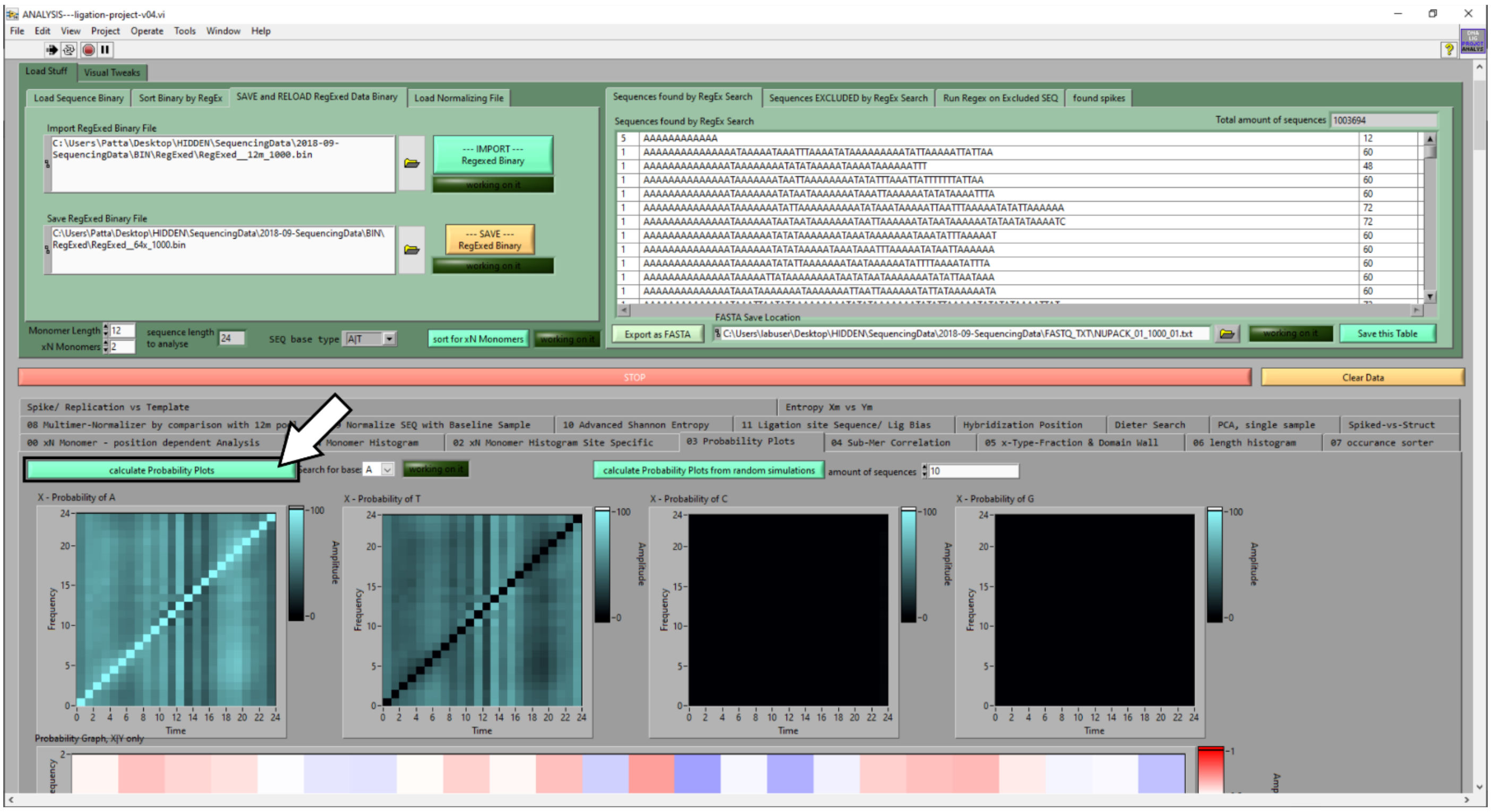

*SI-Fig. 7, **screenshot of the main LabVIEW sequence analysis program:***
*In comparison to most other scripting languages, the front panel of a VI is designed to be a responsive UI. This makes analysis with different parameters and exporting to plotting tools easy.*

SI-Fig. 7 and SI-Fig. 8 show screenshots from parts of the VI front panel and block diagram (and the so called Sub-VIs, which behave like function-calls). The function shown here is run when the button on the front panel is pressed. The function calculates an image, where the probability of all other bases is calculated given a certain selected base (here A). The probability is plotted for all possible positions of the fixed base resulting in a 2D plot of the size oligomer-length squared.

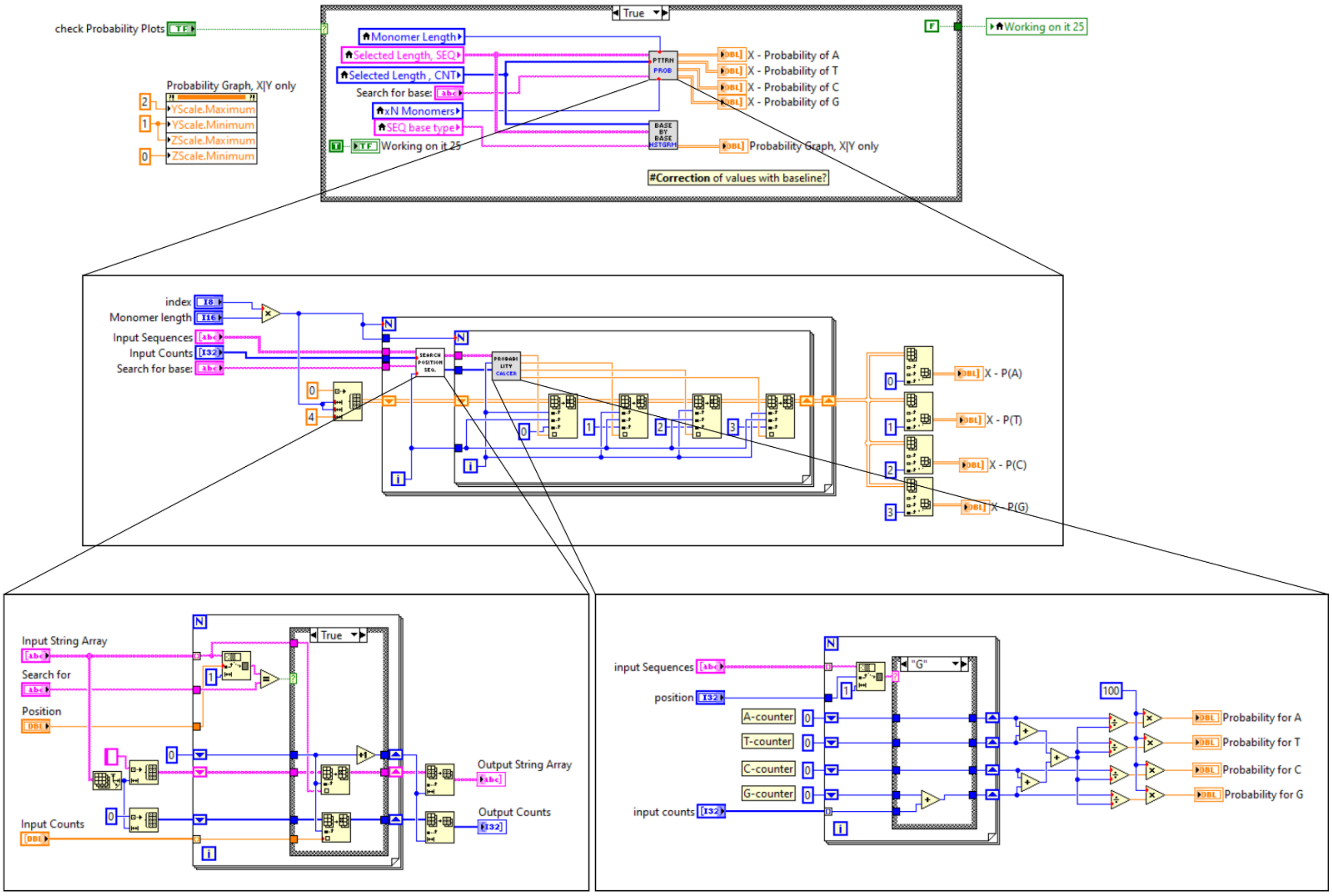

*SI-Fig. 8, **screenshot of the block diagram and Sub-Vis for one example function:***
*The VI-architecture shown here is only one of multiple ways/styles of program in LabVIEW. The top level function is executed when the button in the Front Panel marked in SI-Fig. 7 is pressed. The function then calls several sub-Vis that work like calling a function. In the end the data is printed to graph indicators.*

# 9. Randomness of initial pools

An essential part of the selection experiment is the initial random pool. The randomness of the pool might influence the downstream reaction significantly. Therefore, the randomness of the pool must be considered.

## 9.1. 12mer AT random sequence pool

For a completely random sequence distribution we would expect a binomial-shaped distribution for the A:T content, centered around the 50:50 mark. SI-Fig. 9a shows the 12m AT-only random sequence pool has too many strands with about 60-70 % A and too little strands with 60-80 % T. The lack of poly-A and poly-T sequences is obvious when comparing the distribution to a perfectly random one (simulated, light grey). The simulated curves could be obtained by a simple binomial function, but we opted to calculating a great amount of AT-only random sequence 12mers with the analysis tool and analyzing their distribution with the same A:T-fraction algorithm as the sequence data. The more bases of one kind are found in one strand, the less frequently they are sequenced. Still, in the experiment, there are enough total strands of A-type or T-type for the experiment to work, but the lack of T-type sequences in the monomers might help explain the bias in A-type to T-type in oligomer products.

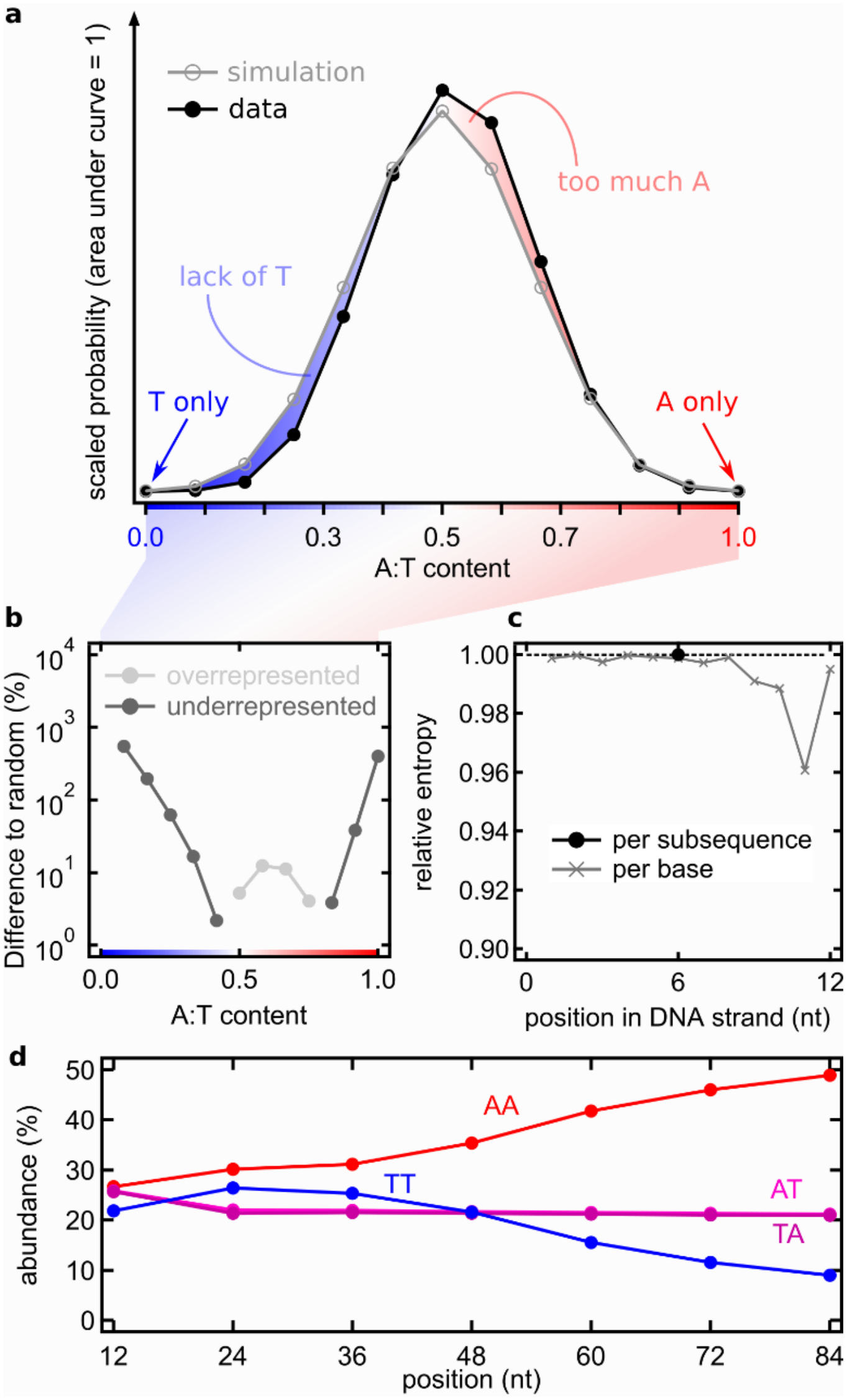

*SI-Fig. 9, **12mer random sequence pool A:T fraction vs simulated random:***
***a*** *Analyzing the A-to-T fraction of the 12mer strands of the remaining pool and comparing them to the corresponding curve of a simulated completely random pool reveals a bias towards A-type sequences. Specifically, ratios of 6:6, 7:5 and 8:4 A:T are overrepresented while 4:8, 3:9 and 2:10 A:T are underrepresented.*
***b*** *The difference of the underrepresented and overrepresented of AT fraction in a) in percent. Poly-A and especially poly-T strands are underrepresented.*
***c*** *The entropy of the ensemble of all strands is not reduced (black circle) due to the large amount of random sequence strands with about 50:50 % A:T-ratio but for the second to last base there is a drop in entropy. This is comparable to Fig. 3a in the main manuscript.*
***d*** *Plotting the abundance of all four possible 2 nt long submotives AA, AT, TA and TT as a function of product length shows, that for the initial pool AA, AT and TA submotives are about equal in abundance, while TT is underrepresented. For longer strands the TT submotives become less abundant. The total amount of A-type sequences is growing, as the increase of the AA submotive to almost 50 % of all 2 nt submotives indicates.*

In the entropy reduction (SI-Fig. 9c), there is only a minimal reduction compared to a perfectly random distribution, if calculated for the entire strand. The analysis for each position shows a lower entropy for the second to last base position. Also, in Fig. 1b in the main manuscript, there is a distinct band in the base probability graph for the 12mers for base A.

Overall, we sequenced 4067 of the 4096 possible submotives (99.29 % coverage). The bias towards more A in the 12mer random sequence pool is probably no artefact, as long oligomers of A-type are significantly more pronounced than T-type sequences. As mentioned in the manuscript, an initial imbalance in A-T content is enhanced in every further elongation of a oligomer strand due to the templated ligation reaction.

Analyzing the abundance of all possible 12mer submotives is a difficult task, and it is even more difficult to visualize the results. In the main manuscript we show that reducing the length of analyzed submotives to a length of 6 nt reduces the amount of motives to $2^6$=64 while retaining almost all local sequence-motive properties. This can also be done for very short submotives of length 2 nt, to analyze an initial or gradual bias in the pool or resulting oligomers.

Therefore, we compare all possible motives AA, AT, TA and TT in SI-Fig. 9d: For the initial 12mers there is a lack of TT motives, while the other three motives AA, AT and TA have about the same abundance. There might have been a small difference in the base attachment probability of a DNA strand during DNA synthesis and is likely the reason for the base-composition bias in the initial 12mers (Fig. 2c). For 24mers, the probability of AT and TA decreases while the probability of AA and TT increase. For longer oligomers the TT- motives get less abundant, while AA gets more abundant. AT and TA motives, mainly responsible for the ligation site A-T alternating sequence patterns stay at about 21 % abundance.

## 9.2. GC random sequence pool

DNA made from bases A and T only has a lower melting temperature compared to ATGC or GC-only DNA because of the weaker Watson-Crick basepairing of A and T, as described in section 6. Nevertheless, the experiment works for GC-only samples as well. The sequences space is $2^{12}$=4096 different sequences (and $2^{10}$=1024 for 10mer monomers).

GC-only samples were subjected to similar experimental conditions, library preparation and illumina sequencing as the AT-only samples. GC-only experiments and subsequent sequencing typically yield significantly less reads and especially less reads for long oligomers, as seen in SI-Fig. 10g and h.

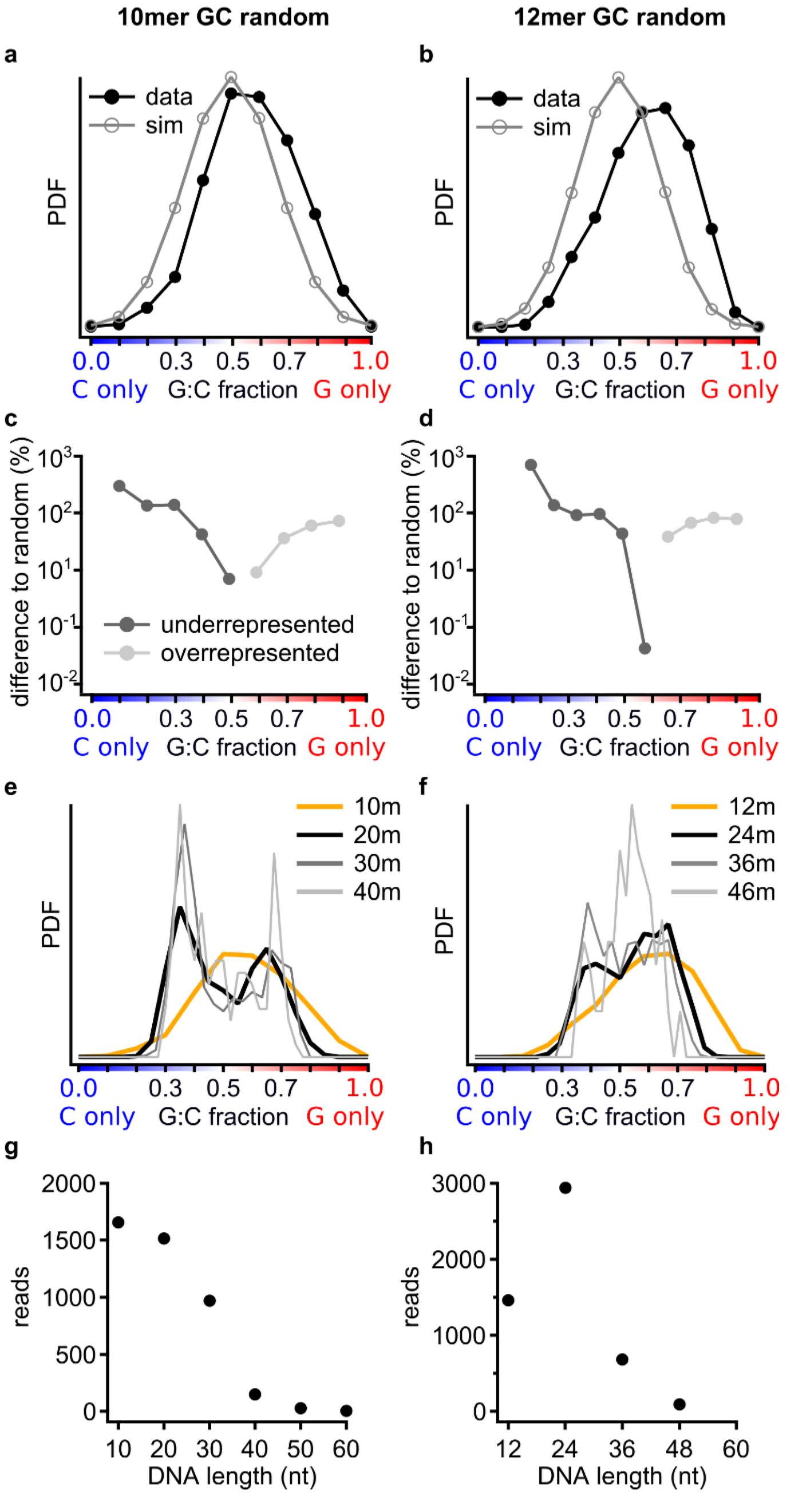

*SI-Fig. 10, **GC-only random 10mer and 12mer samples:***
***a, b*** *For 10mer and 12mer monomers there is a shift towards too much G.*
***c, d*** *G-type sequences are strongly over and C-type sequences strongly underrepresented in both, 10mer and 12mer GC random sequence pools.*

***e*** *10mer GC-only samples start again from an about Gaussian shaped GC-fraction distribution with the before mentioned shift towards G. 20mers and 30mers show the distinct emergence of a bimodal distribution – a G-type and C-type.*
***f*** *12mer GC-only samples only show a bimodal distribution for 24mers. For longer oligomers the amount of analyzed sequences is low and there seems to be no clear shape.*
***g, h*** *Length histograms show only a small amount of reads, although the experimental conditions and library preparation are similar to AT-only samples.*

As shown in SI-Fig. 10, the GC-fraction of 10mer and 12mer samples are comparable to AT-only samples: There is a shift towards too much G, but it is significantly more pronounced than the shift towards A in AT-only. The difference to a completely random distribution is in both cases larger. Additionally, the symmetry of underrepresented poly-A and poly-T sequences (as in SI-Fig. 9b) is not seen for GC. C-type strands are underrepresented, while G-type strands are overrepresented in 10mer and 12mer GC-only "monomers".

The PDF of GC-fractions of different oligomer-lengths shows a bimodal distribution for the 10mer sample. For the 12mer sample, only the 24mer seems to show a bimodal distribution. For longer oligomers the analysis is inconclusive as the amount of analyzed sequences is substantially smaller than for AT-only samples due to a very low readout success-rate in GC-only illumina sequencing.

Interestingly, for the 10mer samples, the bimodal distribution seems to favor C-type sequences, although the initial 12mers are predominantly G-type sequences. DNA with a lot of bases G are often referred to as "sticky". Guanine domains tend to form duplexes with other strands or fold on themselves[3]. With this additional binding mechanism absent in AT-only DNA the G:C-fraction might favor longer C-type sequences, because G-type sequences stick together or to themselves and are thus not taking part in templated ligation reactions.

### 9.3. Error estimation in random sequence pool sequencing

In contrast to sequencing data of well-defined strand elongation experiments, where it is possible to already include the primer, barcode and possibly the binding section to adhere the strand in the flow cell of the illumina sequencer, this is not possible here. The short 12mer DNA strands need be able to be ligated on both 5' and 3' end without a long tail of additional sequence for later characterization. Therefore, we utilize, as described in the methods section of the main manuscript, a library preparation kit to enzymatically attach a primer and barcodes to our sample strands of different lengths. This attachment might be biased towards certain sequence motives. As there can't be a "real" reference sample because of the randomness of the pool, we can only estimate and search for systematic biases in our results. There are several results that point towards different errors:

- The comparison of 2 nt motives in strands of all lengths (SI-Fig. 9d) shows, that motives AA, AT, TA are comparable in abundance, while TT is underrepresented. This might hint towards a bias in strand synthesis. In longer strands the abundance of AT and TA is about constant while the oligomer products are segregated into A-type and T-type strands, with T-types being less abundant due to the initial asymmetry.
- Although there might be bias in attaching the random sequence CT-tail to 12mers and their oligomer products during library preparation, this enzymatic elongation (possibly done by a terminal transferase, unclear because the manufacturer doesn't provide further information) is likely only biased by the few bases close to the 3'-end of the strand. SI-Fig. 9c suggests a very small bias in the entropy for bases in positions 9, 10, 11 and 12 that might stem from either DNA synthesis or sequencing. Despite that, most analysis of sequencing data performed here is done on the entire strand or the only the center subsequences, that are predominantly far enough away from the 3'-end to be bias from the selection. In graphs like Fig. 2, 3 and 4 of the main manuscript, results show great reproducibility over different lengths, and the results fit the theoretical models well (that don't have a bias for sequencing or synthesis artifacts).

Both points support the assumption, that the library preparation kit is well suited to prepare random sequence stands for NGS without introducing strong biases. The asymmetry of the pool and underrepresented sequence motives might account for greater uncertainties in comparison to an error due to library preparation or sequencing.

## 10. Entropy Reduction

The entropy calculation used here is done as described by Shannon[4] and adapted as by Derr *et al.*[5]:

1) $H_k(s) = -\sum_i p_i \log_2(p_i)$

with the entropy $H_k(s)$ of a sequence $s$ of length $L$ with $i$ as the index of a unique substring of length $k$. The frequency of the $i$th $k$-mer in $s$ is called $p_i$. For the entropy of single base positions, the length of the substring is 1, for the subsequences it is 12. For the plot in Fig. 2a we calculated the entropy of a large set ($10^7$) of randomly generated AT-sequences with the same algorithm and divided it by the value calculated for the sequencing data. A value of 1 then corresponds to no reduction in sequence entropy in comparison to an ensemble of random sequences. A value close to 0 represent the case of a very low sequence diversity with only a handful of sequences in all strands. For the entropy reduction in single positions the y-axis is scaled by a factor of 12 for easy comparison with the 12mer subsequences.

## 11. Pearson Correlation Coefficient

Comparing two populations with the same elements but different distributions can be done with the sample Pearson Correlation Coefficient (sPCC). On the x-axis the elements of one distribution are plotted, sorted from smallest probability to highest probability. The same is done for the second population but on the y-axis. If the populations correlate, elements that are common in one, are also common in the other, the same is true for uncommon elements. The resulting plot can be described by a linear function. If the populations don't correlate, the resulting distribution has no particular shape. SI-Fig. 11b shows all possible sPCC plots for the comparison of all 36mers to all 48mers and all subsequence positions. The first, the last and the two center subsequences have high correlations, while the first compared to last have very low correlations. SI-Fig. 11b also plots the A:T-ratio: the highly correlating start sequences have two distinct populations for the A-types (blue) and the T-types (red) with different slopes. There is a clear trend towards more A-type sequences in longer oligomers, indicated by the flatter in slope. The plot in SI-Fig. 11a is the color representation of the sPCC values calculated from the linearity of the correlation plots.

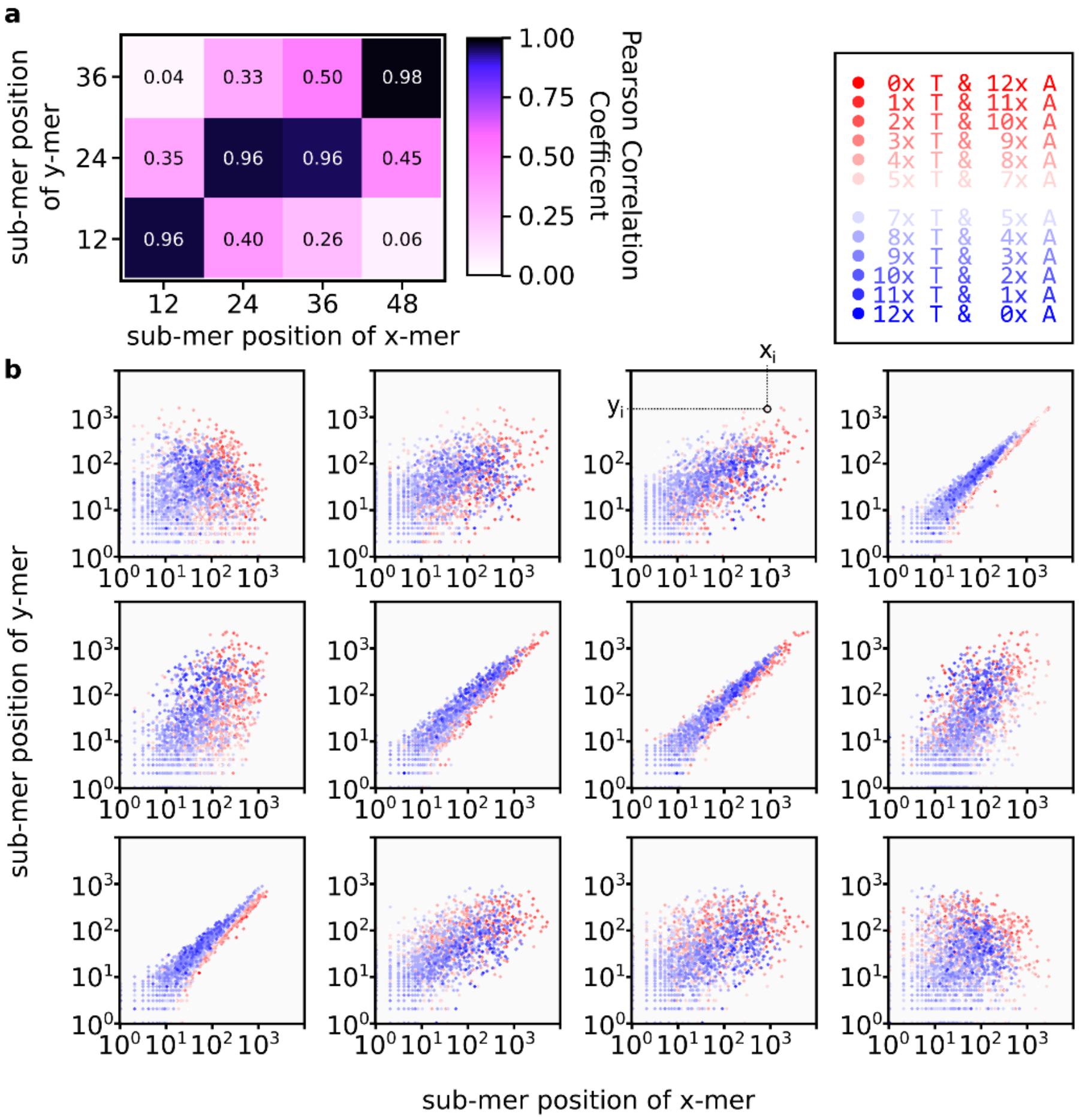

*SI-Fig. 11, **sample Pearson Correlation Coefficient for 48mers vs 36mers:***
***a*** *Color-coded sPCC from the sub-plots shown in b.*
***b*** *sPCC plots for all possible combinations of 12mer subsequences for the comparison of 36mer and 48mers. The color codes the A:T fraction for each sequence. In the second subfigure from the top right, points $x_i$ and $y_i$ from equation (3) of the main manuscript are shown.*

Comparing the 36mer A-type sequences to the 36mer T-type sequence reveals a slight negative correlation, as shown in SI-Fig. 12a. Strands with a lot of A are uncommon in T-types and *vice versa*. But comparing the 36mer A-types with the reverse complement of the 36mer T-types reveals an already known pattern. T-type sequences are just the reverse complement of A-type sequences. This is no surprise, as both groups can function as the template for one another.

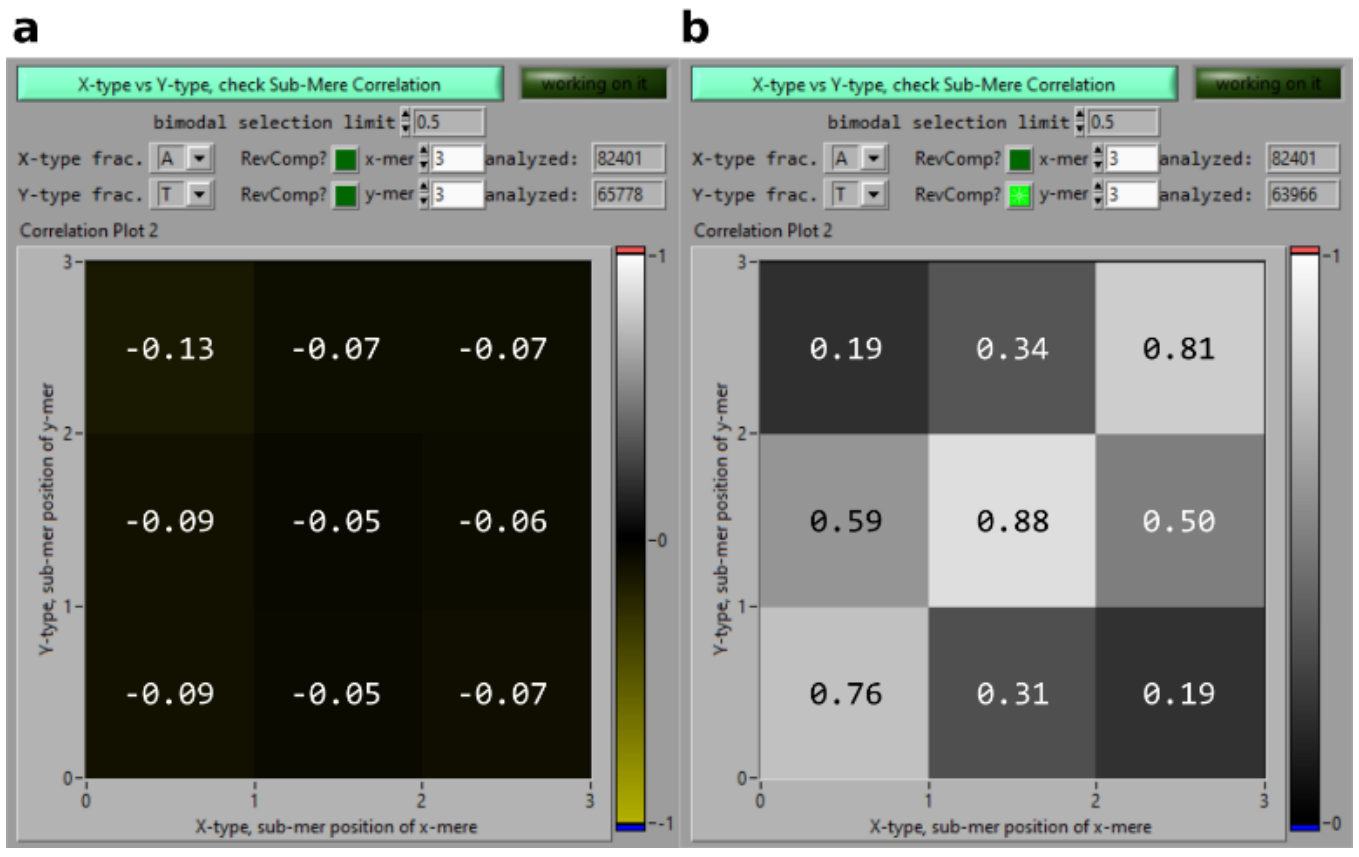

*SI-Fig. 12,* ***Pearson Correlation Graph for 36mer A-type vs 36mer reverse complement T-type strands:***
*Screenshot of the LabVIEW analysis tool.*
***a*** *Comparing 36mer A-types and 36mer T-types reveals no correlation.*
***b*** *Comparing the reverse complement of the T-type sequences instead gives a very similar pattern as already known for the comparison of different lengths.*

## 12. Domain Wall

Starting with the 24mer products, oligomers can be categorized as either A-type or T-type sequences with A:T ratios of about 70:30 or 30:70. But for long oligomers like 72mers the originally bimodal distribution starts to has additional small peaks around A:T fractions of 0.45 and 0.55. SI-Fig. 13 shows the most common subsequence-sequences. Here, every 12mer subsequence is analyzed for its A:T content, if a strand has more than 50 % A it's called "a", for more than 50 % T it's called "t". So a 36mer strand like "AATTAATAAATA-TAATTAAATTTT-AAAAAAATAATA" is noted as "ata". This results in a histogram of 6-sequences long representation of a strand.

As expected, the majority of strands is either completely A-type ("aaaaaa") or completely T-type ("tttttt"). The second most common group is strands with a single subsequence not matching the others, like in an A-type strand a subsequence-sequence like "aaaata".

The small additional peaks mark strands with significant read counts that do not match the first two groups and are either 1:1 A- and T-type ("aaattt" and "tttaaa") marked in light grey, or they consist of 2 non-matching subsequences, like "ttaaaa" marked in dark grey. Interestingly, the non-matching subsequences are all consecutive: "aaaatt", "aatttt", "ttaaaa" and "ttttaa", even "taaaat". This is in agreement with the possible self-folding mechanism and the templated ligation mechanism in the experiment discussed in the main manuscript. This particular effect shown here is best described as a domain wall: if a sequence has a certain appearance, like "aaaa" there is a higher chance to attach "aa" than "at", which in turn has a higher probability than "tt".

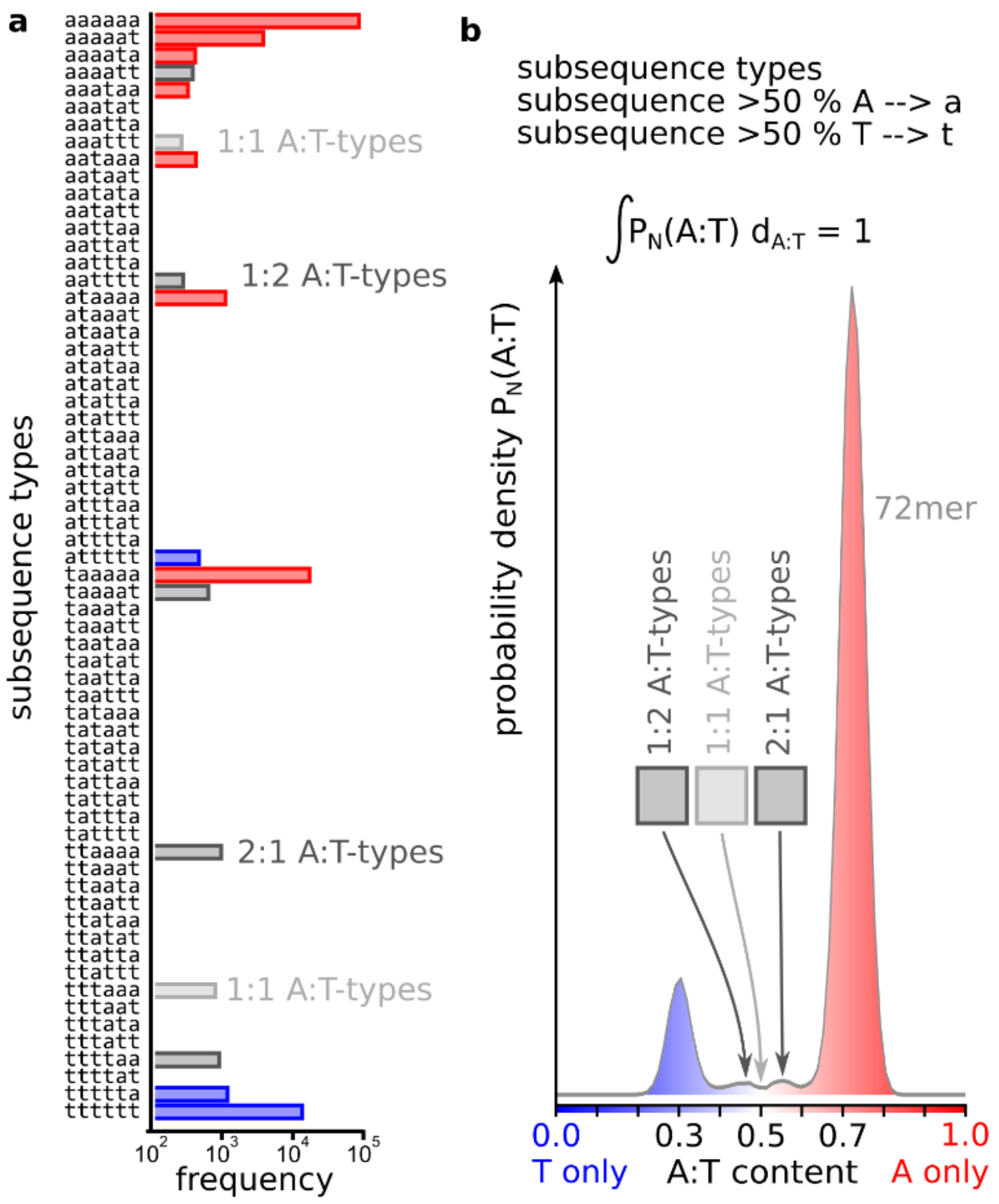

*SI-Fig. 13,* ***Most common subsequence types and comparison with AT-composition graph:***
***a*** *Noting A-type and T-type subsequences in a 72mer as "a" and "t" shows, that not only the entire strand is made from 70 % of A or T, but so is each subsequence. Rare subsequences include aaattt and tttaaa that might fold on themselves and sequences with two non-matching subsequences, like aatttt. In those strands, the non-matching subsequences are predominantly consecutive.*
***b*** *The groups of subsequence-sequence motives give rise to additional peaks in the originally bimodal A:T-fraction graph shown in the main manuscript.*

# 13. Correlation of Ligation Sites, Inter-Ligation Sites, Strand Start and Strand End

The ligation site and the inter-ligation sites show distinctly different sequence patterns, as discussed in the main manuscript Fig. 4b. But the Analysis of the sPCC shows, there should also be a clear difference between the start and the end of the oligomers. SI-Fig. 14 shows the distribution of all 64 possible 6 nt long submotives in 36mer oligomers. On the bottom, the start and end strands are plotted. They do have certain similarities like a peak for the sequence AAATTT, but are overall clearly different. The same holds for the ligation site and the inter ligation sites and both are again different, compared to the start and end sequences. As suggested by the sPCC the combination of different submotives describes where they can be found in a oligomer.

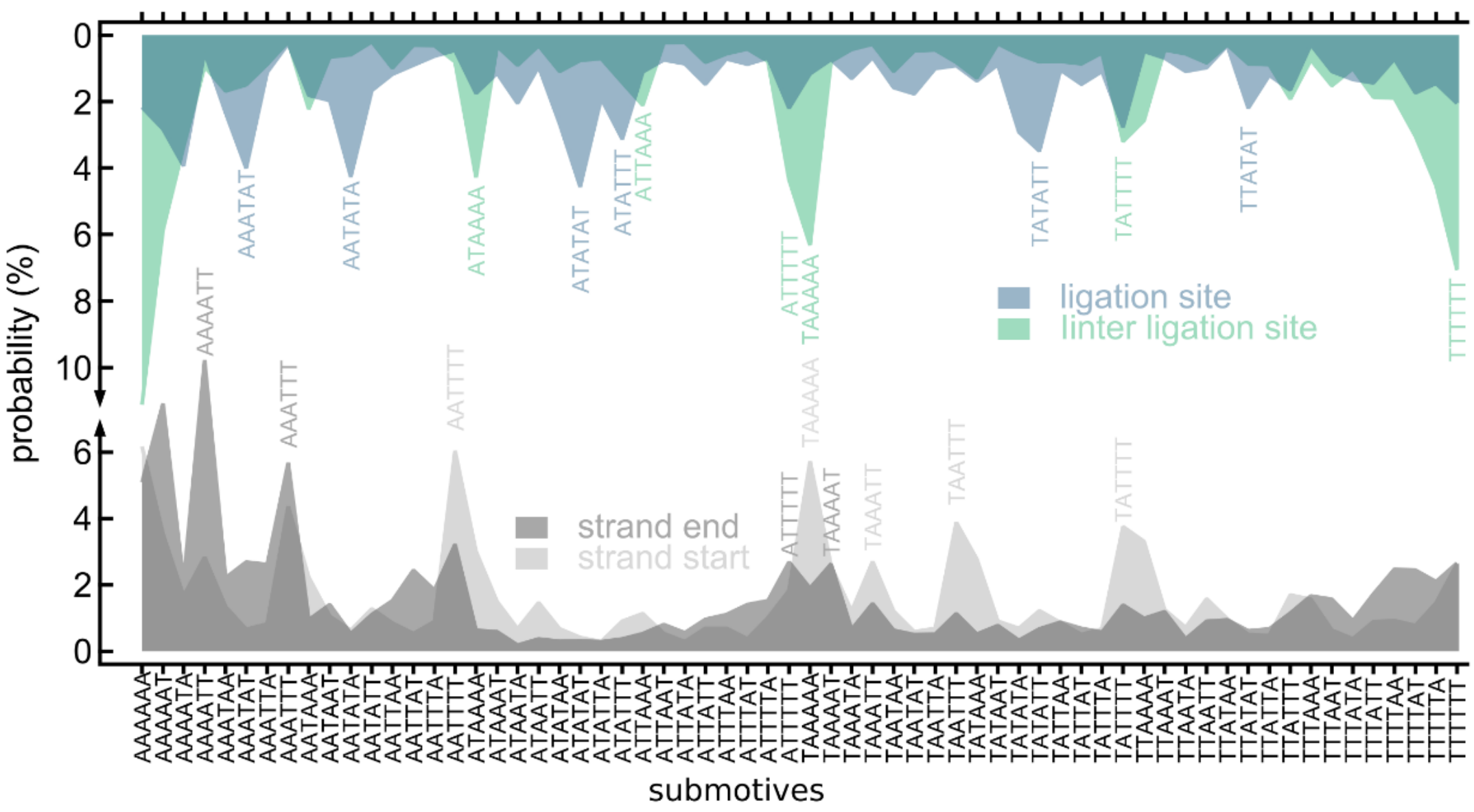

*SI-Fig. 14,* ***6 nt long submotives on 36mer reaction products, analyzed in groups of strand position:***
*Plotting the probability to find a submotive in the region of the ligation site (blue) and in between ligation sites (green) shows, they have different distributions. This is known from Fig. 4b of the main manuscript. The graph suggests, that there is no clear correlation between start- (light grey) and end-of-strand (dark grey) submotives. Interestingly, there is almost no correlation between any of those four groups of submotives.*

## 14. Z-score landscape

The z-score is used when comparing differently scaled and distributed normal distribution datasets. The z-score shifts and normalizes the dataset by the mean value of the distribution. Data points above the mean of the dataset are given in positive values, data points smaller than the dataset mean are represented by negative values. Equation 1) gives the mathematical representation of the z-score of a sample with index *i*, *j*:

$$2)\quad Z_{ij} = \frac{x-\mu}{\sigma} = \frac{N_{ij}^{\text{observed}} - N_{ij}^{\text{expected}}}{\sqrt{N_{ij}^{\text{expected}}}} \quad \text{with} \quad N_{ij}^{\text{expected}} = \frac{N_i N_j}{N_{\text{total}}}.$$

For the z-score landscape in the main manuscript transitions with either significantly higher or significantly lower than mean probabilities are plotted.

## 15. First strands ligation

At some point in the experiment during temperature cycling the first ligation events occurs. There is no possibility to find the exact conformation for those events; the results might also be included in longer strands. They might also just be precursor needed for the first reactions that is then not rebuild and simply be so rare, that they are not sequences. Therefore, some estimations must be made to get a simple model for such a reaction. SI-Fig. 15 shows a schematic sketch of a likely conformation. Two strands from the 12mer pool are brought together by a third strand, that templates the other two by Watson-Crick basepairing. Such a complex can be ligated by the ligase enzyme. The products are then the connected 12mers, now as a single 24mer ssDNA strand, and the templating 12mer.

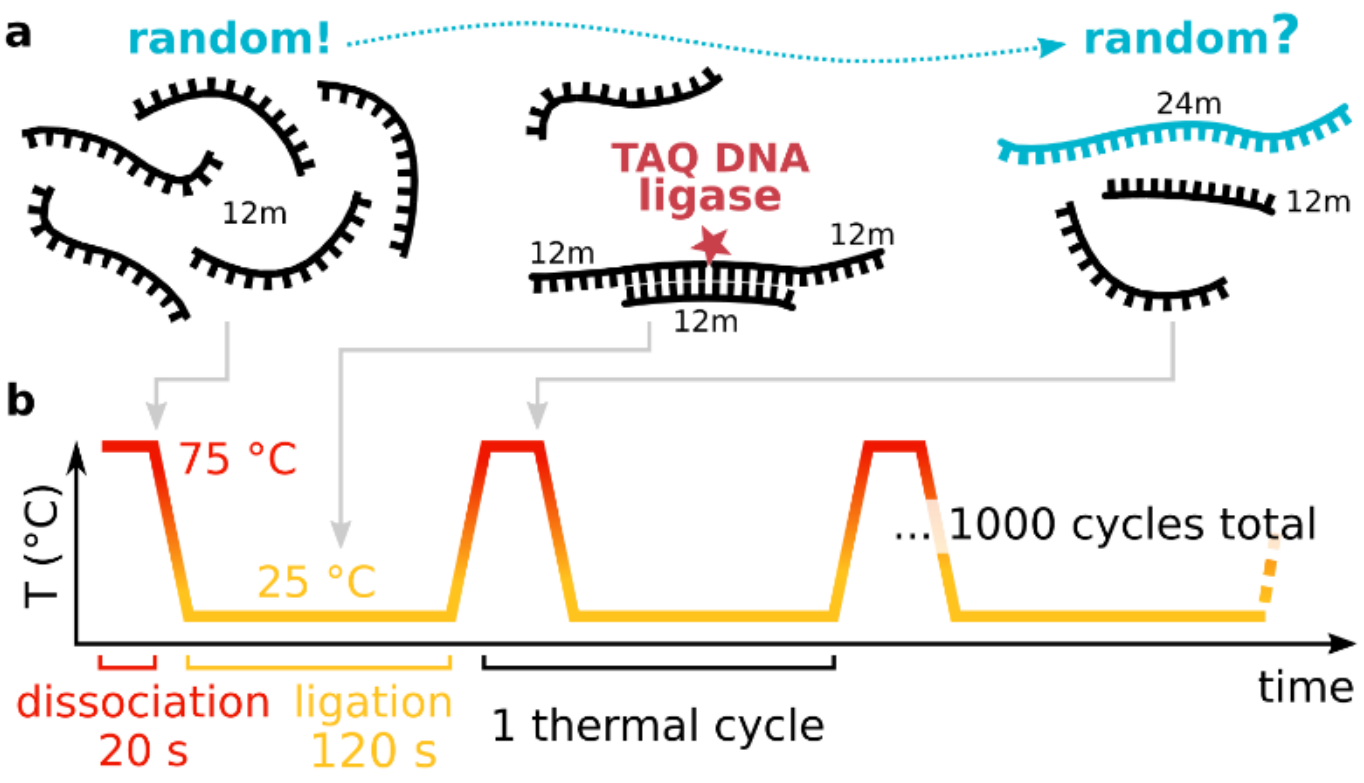

*SI-Fig. 15, **schematic drawing of the hypothesized conformation for the first ligation events:***
***a** The first ligation events must have occurred solely between 12mer monomer strands. As a simple model, those were in a substrate+ substrate +template conformation as shown. Two 12mer strands are brought 3'-end-to-5'start by a templating third strand. The ligase can connect such structures and the result is a 24mer and the templating 12mer.*
***b** Shows the different pool and structure conformations in an exemplarily drawn temperature cycle. At dissociation temperatures, there aren't any dsDNA strands. At the ligation temperature, strands might be in conformations, that can be ligated by the ligase. In subsequent dissociation, those complexes are melted to ssDNA again.*

In this simple model, there is no dynamic aspect included: one might expect to get 24mer sequences as a results for a small amount of temperature cycles already. But as shown in Fig. 1 of the main manuscript, the first substantial bands are 36mer and 48mer, while 24mer sequences seem very scarce. Those bands can hardly be seen by eye and even for higher cycle counts, the resulting band structure suggest a non-trivial growth mode. Therefore, it's not possible to conclude, that 24mers that were sequenced have a close correlation to the first emerging sequences.

## 16. Ligation site shift

In the graph showing the probability for submotives on 72mer junctions (Fig. 4a, b in the main manuscript) there is a cluster of junctions with a higher probability of including poly-A on the ligation junction. This is not expected, as all graphs up until here show a distinct AT-alternating pattern for ligation junctions. This effect might best be described as a hybridization or ligation site shift. With a long single stranded unbound oligomer, there is no reason shorter strands like a monomer should hybridize exactly onto a 12mer subsequence. It might rather hybridize with its center on a ligation site. If this monomer is extended by another strand, the AT-alternating and poly-A, poly-T pattern the product inherits from the template is shifted relative to the 5'-end of the strand. The abundance and length dependence of this effect is further described in the following.

Detecting those shifts is done by analyzing the submotives on ligation junctions. As the analyzed strands are all a multiple of 12 nt in length, the junctions have to be after a multiple of 12 read bases. Filtering oligomer-junction submotives for poly-A and poly-T of lengths of minimum 4 nt and maximum 6 nt in a region of plus and minus three bases from the junction produces the data shown in SI-Fig. 16.

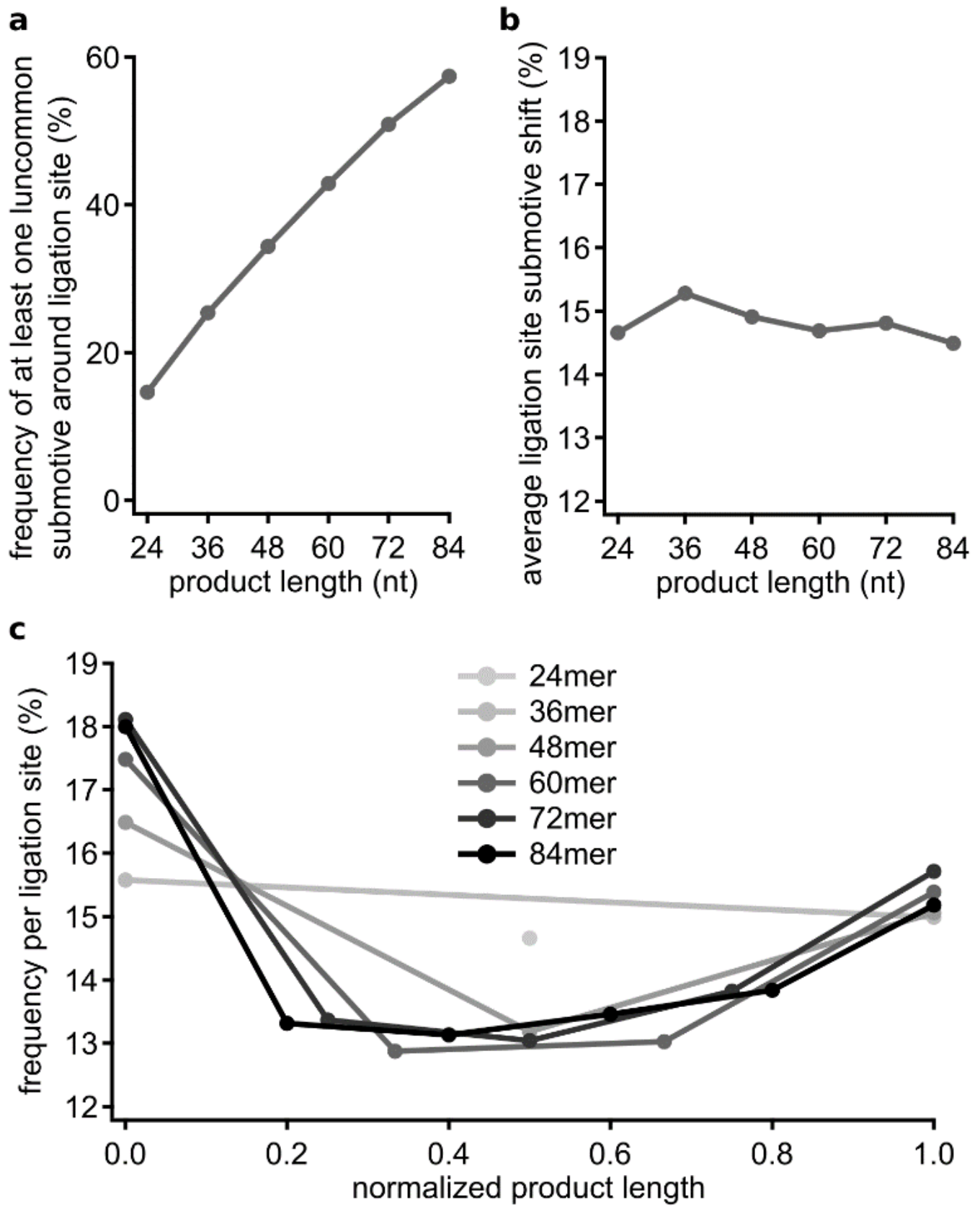

*SI-Fig. 16, **Ligation site shift:***
***a** Frequency of finding at least one uncommon submotive at a ligation site per strand. Longer strands have more junctions and an about linear increase for the probability of having at least one uncommon submotive around at least one ligation site.*
***b** The average probability of an uncommon submotive per ligation site as a function of the product length is about 15 % for all lengths.*
***c** The ligation site submotive shift shown in b is not constant per length. The probability to find an uncommon submotive on a ligation site is larger on the outer ligation sites.*

SI-Fig. 16a shows the frequency of finding at least one of those uncommon poly-A or poly-T submotives on at least one ligation junction. With a linear increase, longer strands like 84mers have a chance of about 60 % to include at least one ligation site shift. In SI-Fig. 16b the abundance per junction is analyzed, and as expected from a the average ligation site shift probability is about constant in all oligomers and about 15 %. But SI-Fig. 16c shows, this frequency is not the same for all junctions. The junctions in the oligomer center have a lower probability of including a poly-A or poly-T submotive compared to the outer strands. This can be seen for all oligomer lengths.

After all, Fig. 1b and Fig. 4a, b of the main manuscript suggest, that overall the ligation site is most common after a multiple of 12 bases.

# 17. x8 experiment, designed and selected pools of eight sequences

## 17.1. “Replicator” design

For the comparison of the most common selected sequences of the AT-random sequence 12mer pool we build a set of eight sequences designed to elongate. As described in section 15, we chose strands that are able to form three strand complexes, with one templating strand and two substrates. The sequences are then made from regions of alternating bases and poly-bases: All strands have either two poly-base parts and one alternating bases part, or *vice versa*. For every strand there is also its reverse complement strand, as shown in SI-Fig. 17.

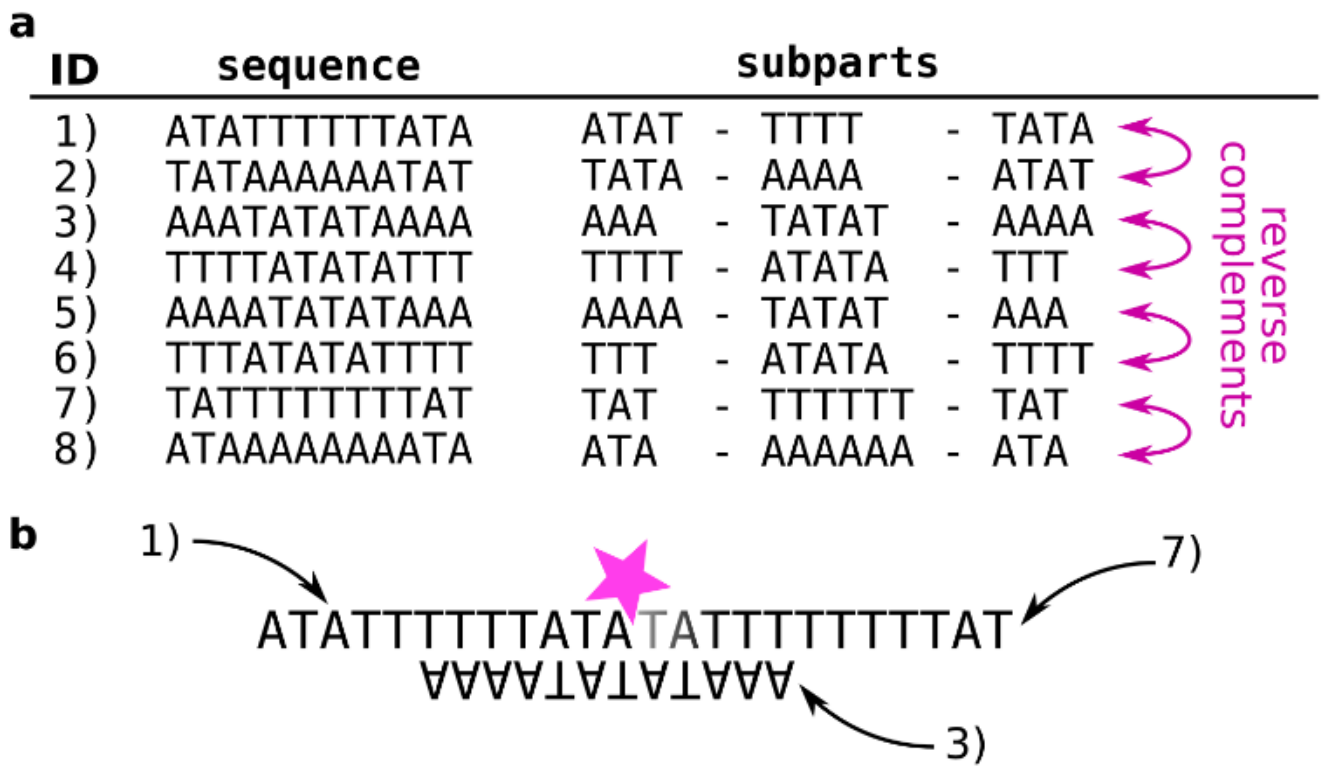

*SI-Fig. 17, **x8 "Replicator" sequence design:***
***a** Sequences and subsequence parts of the Replicator pool. There is either alternating or poly base sections, always in a 1-2-1 or 2-1-2 conformation. The stretches differ in length in order to form properly ligatable dsDNA conformations.*
***b** An example for a presumed ligation construct in early temperature cycles. Strands with IDs 1) and 7) act as the substrate while being templated by strand 3).*

### 17.2. x8 experiment after 50 temperature cycles

The experiment comparing three pools of eight sequences each described in the main manuscript was done for 50 and for 200 temperature cycles. Fig. 5c shows the PAGE gels and Fig. 5d the concentration quantification for 200 temperature cycles. SI-Fig. 18b shows the concentration estimation for 50 temperature cycles. The inset shows the concentrations for the 12mer band. The Random sample depletes the pool the least, by not even 10 %, while the Network depletes the pool the furthest. The oligomer product concentrations are also the lowest for the Random sample, while the Replicator sample and the Network sample have both about exponentially decaying product concentrations over length.

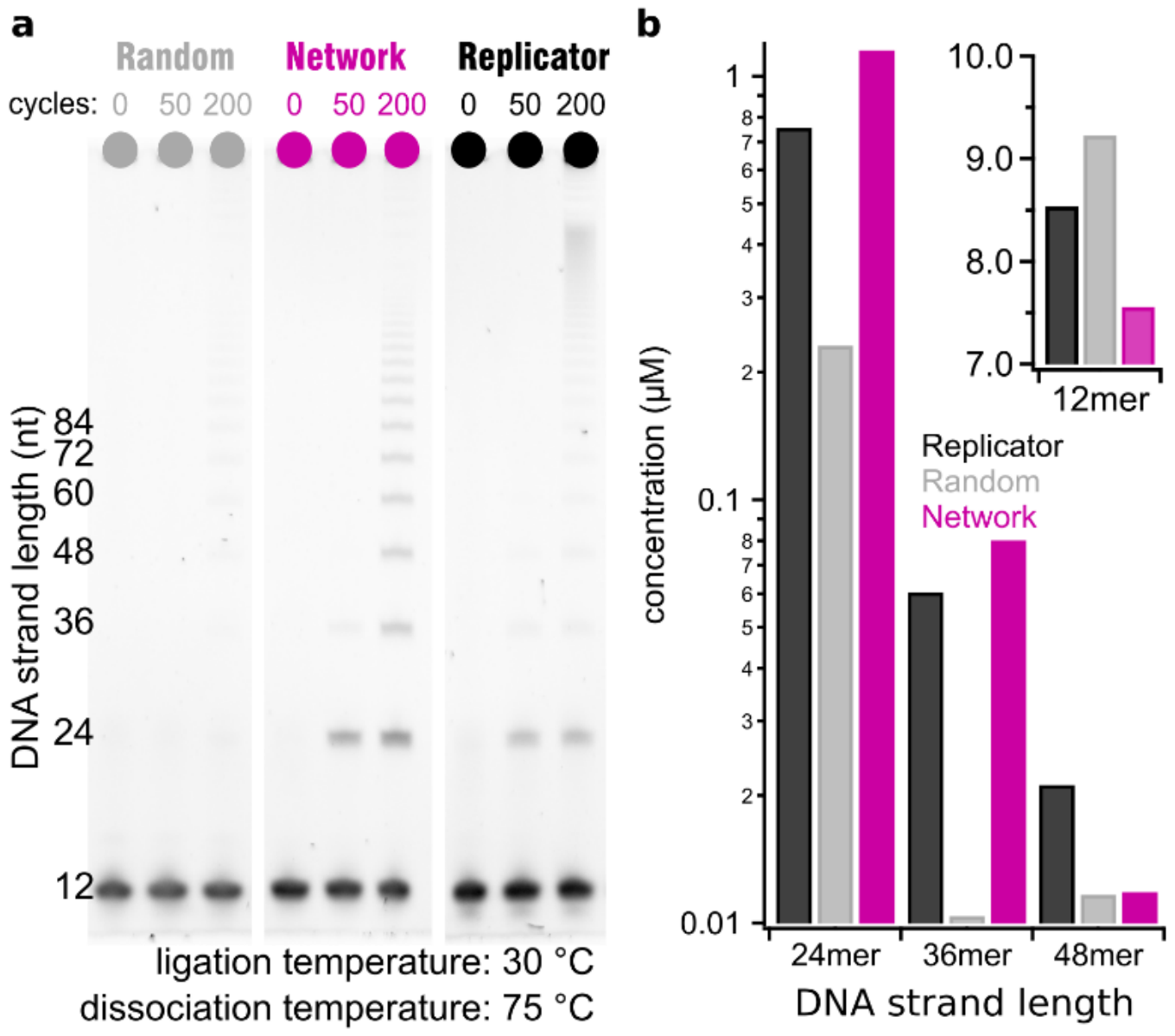

*SI-Fig. 18, **concentration quantification for x8 pool comparison and 50 temperature cycles:***
***a** Graph similar to Fig. 4c in the main manuscript showing a PAGE gel of the x8 pools comparison.*
***b** Concentration quantification of the PAGE gel in a, drawn as a bar plot showing the concentrations after 50 temperature cycles.*

## 18. A:T-composition and kinetic simulations

### 18.1. Composition and hairpin formation

One striking evidence of the selection in the system is the formation of two subpopulations of oligomers: A-type and T-type. A likely origin of this behavior is that the templating ability of a given sequence would be substantially reduced by the formation of internal hairpins and other secondary structures. The same is true for the substrate chains. An easy way to suppress the formation of hairpins is by introducing a composition bias in a sequence. Consider a sequence of length $N$, which is random but whose composition $p$ (i.e. the fraction of 'A'-bases among all nucleotides) is different from 0.5. How likely is it to find an internal hairpin of length $l_0$, i.e. two non-overlapping mutually complementary regions of length $l_0$, within that sequence? The probability of any two given bases in the sequence to be complementary is $2p(1-p)$, which yields the probability $[2p(1-p)]^{l_0}$ for two given segments of length $l_0$ to be complementary. The number of ways in which two such non-overlapping segments can be chosen on a sequence of length $N$ is $(N-2l_0)^2/2$ (assuming large $N$). By requiring the expected number of hairpins of length $l_0$ to be 1, we obtain an equation that relates the sequence length $N$ to the length of a typical maximum hairpin within it:

3) $N = 2l_0 + \sqrt{2}[2p(1-p)]^{-l_0/2}$

If the sequences generated by our autocatalytic reaction were completely random, subject only to the constraint of the fixed mean composition for both T-type and A-type chains, we would obtain an ensemble of sequences that maximizes entropy under that constraint. The result of such maximization is a regular Gibbs-Boltzmann distribution, with abundances of individual sequences proportional to $\mathrm{e}^{\lambda p}$(see following section). Note that the assumption of maximum entropy is excessive. We know that the entropy of the generated oligomer pool is well reduced beyond a simple compositional bias. However, if all other selection is not correlated with the value of $p$, the maximum entropy ensemble gives a correct probability distribution function (PDF), $f(p)$. In order to obtain one, we need to account for the "Boltzmann factor", and the total number of sequences with a given composition. The latter follows a regular unbiased binomial distribution (sequence of length $N$ is statistically equivalent to $N$ sequential tosses of a coin), and can be well approximated by a Gaussian curve $\sim\mathrm{e}^{-2N(p-1/2)^2}$. This gives the following general result for a compositional PDF in chains of length $N$:

4) $P(x) = \sqrt{\frac{2N}{\pi}}\left(\frac{\beta^{\left(\frac{N}{12}-2\right)}\mathrm{e}^{-Nx_0x}+\mathrm{e}^{Nx_0x}}{\beta^{\left(\frac{N}{12}-2\right)}+1}\right)\mathrm{e}^{-2N(x^2+x_0^2)}$

Here $x = p - 1/2$ , the mean compositions of A-type and T-type subpopulations are $1/2 \pm x_0$. The empirical parameter $\beta$ accounts for observed uneven abundances of A- and T-type oligomers starting with $N$=36. The exponential enhancement of the contrast between the two subpopulations is a direct consequence of the exponential chain length distribution.

## 18.2. Kinetic model

We adapt the model developed in reference[6] to include effects of a hairpin formation which reduces the activity of the reaction of both template and substrate strains. Specifically, we describe the autocatalytic formation of 24mers from the respective pairs of 12mers by the following equation:

5) $\dot{d}_{ij} = \lambda\left(\alpha_{j^*i^*}d_{j^*i^*}\right)(\alpha_i l_i)(\alpha_j r_j)$

Here $d_{ij}$, $l_i$ and $r_j$ are concentrations of a specific 24mer and its constituent "left" and "right" 12mers. $d_{j^*i^*}$ is the concentration of its complementary 24mer which within this model acts as a sole template for $ij$. $\lambda$ is the ligation-rate which was assumed to be sequence independent for simplicity. Coefficients $\alpha$ are activities of the respective DNA fragments (12mers $i$ and $j$, and 24-mer $i^*j^*$) that depend on the length of the longest internal hairpin $l_s$ for a sequence $s$:

6) $\alpha_s = \frac{1}{1+e^{-(G_0+\Delta G l_s)/\mathrm{kT}}}$

Here $\Delta G$ is hybridization free energy per base ($\Delta G \approx 1.5\ kT$ for random AT-based sequences), and $G_0$ is a threshold free energy that accounts for termination of the hybridized region (about 1.5 $kT$ at each side), and for the loop formation. We assume $G_0 \approx 6\ kT$.

For each of the $2^{12}$ 12mers and $2^{24}$ 24mers, we have determined the longest hairpin length, and the corresponding activity factor. We then solved the set of equations numerically, and observed the development of the bimodal composition profile, indicating formation of A-type and T-type subpopulations within 24mers.

This kinetic model only accounts for templating of 24mers by other 24mers. It is therefore only applicable to relatively early stages of the experiment. Once longer oligomers start appearing, one needs to account for sequence-dependent suppression of activity due to formation of hairpins within those longer chains. As discussed above, the average composition can be used as a proxy to find the length of the longest internal hairpin within a sequence. As a result, the overall composition PDF is well described by the minimalistic, maximum entropy model presented above.

# 19. Network-families resemblance

Fig. 5a and b in the main manuscript show 12mer subsequences common in oligomers longer than 48 nt as a de Bruijn network graph, that depicts each sequence as a node and sequences that follow each other are connected by an edge. Nodes and edges are scaled with the abundance of the sequence and the connection.

As stated in Fig. 4 and SI-Fig. 12 A-type and T-type sequences are mostly the reverse complement of each other. In the eight most common A-type and T-type sequences, there are four sequences of which the reverse complement can be found in the other respective group (reverse complement dark pink, other four most common sequences light pink color). The network of the most common sequences is not entirely connected, but consists of several "families" of sequences. Those families can themselves be a very intricate network, or just contain one sequences that tends to follow itself.

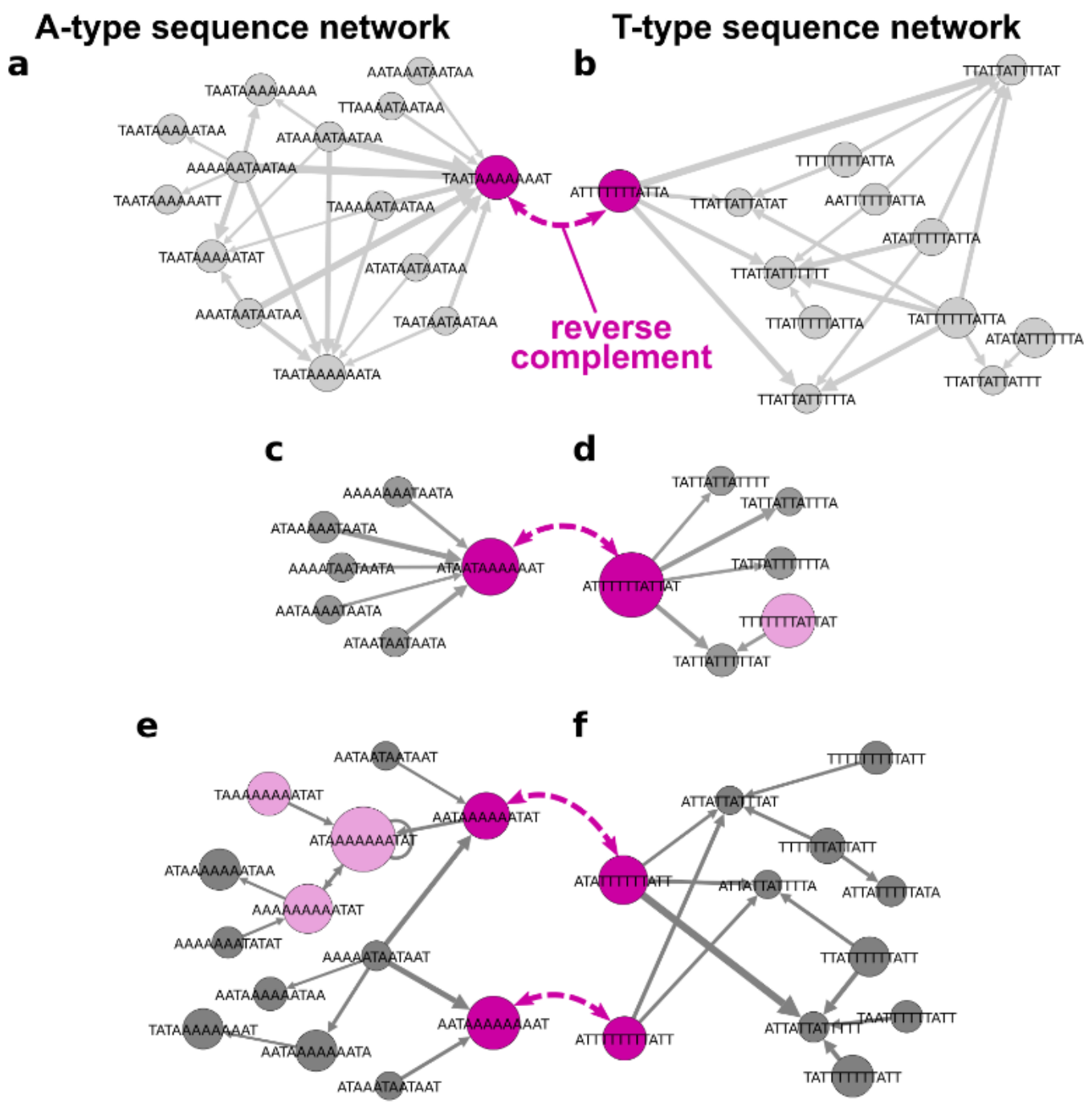

*SI-Fig. 19, **Network families show similar network structures:***
***a & b** Both families are intricate networks, that have multiple internal connections as well as several start and end sequences.*
***c & d** Both networks are very directed, with the A-type sequence ending in one specific sequence, and the T-type sequence, starting with a specific sequence.*
***e & f** The two families both have two sequences that are found in the most common and reverse complement sequences of the network. The sequences are connected in the network by one (A-type) and two (T-type) intermediate sequences.*

SI-Fig. 19 shows the six network families with the four most common and reverse complement sequences. The families of A-type and T-type with the respective common reverse complement sequence each have similar forms, that are distinctly different for the three groups. SI-Fig. 19 a, b are a network with multiple internal connections and several start- and end-sequences. In contrast, for SI-Fig. 19 c, d the networks are very directional. The A-type sequence is the end-sequence for all connected sequences, while the T-type is the start for all connected sequences. And in both cases, the attached sequences have no relevant edges connecting them among themselves. In SI-Fig. 19 e, f the families each contain two sequences that have a reverse complement in the other respective family. In the A-type there is one, in the T-type there are two intermediate sequences connecting them in the network graph. In both cases, there are no edges that would suggest commonly finding both sequences in the same oligomer strand.

Despite the differences in the form of the families, the new initial pool made from the most common four A-type and four T-type sequences has a high oligomer production rate and compared to the total amount of oligomer products a remarkably similar sPCC matrix. This is despite the networks explicitly not taking the first and last subsequence in oligomer products into account – as those are distinctly different to the center subsequences (discussed above).

## 20. List of DNA used

| List of DNA used | | | |
|---|---|---|---|
| **Name** | **Length (nt)** | **Sequence (5' to 3')** | **Modification** |
| AT-random_12m | 12 | WWWWWWWWWWWW | 5'-POH |
| GC-random_12m | 12 | SSSSSSSSSSSS | 5'-POH |
| GC-random_10m | 10 | SSSSSSSSSS | 5'-POH |
| AT_x8_Replicator_01 | 12 | ATATTTTTTATA | 5'-POH |
| AT_x8_Replicator_02 | 12 | TATAAAAAATAT | 5'-POH |
| AT_x8_Replicator_03 | 12 | AAATATATAAAA | 5'-POH |
| AT_x8_Replicator_04 | 12 | TTTTATATATTT | 5'-POH |
| AT_x8_Replicator_05 | 12 | AAAATATATAAA | 5'-POH |
| AT_x8_Replicator_06 | 12 | TTTATATATTTT | 5'-POH |
| AT_x8_Replicator_07 | 12 | TATTTTTTTTAT | 5'-POH |
| AT_x8_Replicator_08 | 12 | ATAAAAAAAATA | 5'-POH |
| AT_x8_Random_01 | 12 | AAAATAAAATAT | 5'-POH |
| AT_x8_Random_02 | 12 | ATAATTAAATAA | 5'-POH |
| AT_x8_Random_03 | 12 | TAAAAATTATTT | 5'-POH |
| AT_x8_Random_04 | 12 | TTAAATTTTATA | 5'-POH |
| AT_x8_Random_05 | 12 | TATTTAATTTTT | 5'-POH |
| AT_x8_Random_06 | 12 | TAAAAATTAATA | 5'-POH |
| AT_x8_Random_07 | 12 | AAAATAATTTAT | 5'-POH |
| AT_x8_Random_08 | 12 | TTATATAAAATA | 5'-POH |
| AT_x8_Network_A-type_01 | 12 | ATAATAAAAAAT | 5'-POH |
| AT_x8_Network_A-type_02 | 12 | AATAAAAAAAAT | 5'-POH |
| AT_x8_Network_A-type_03 | 12 | AATAAAAAATAT | 5'-POH |
| AT_x8_Network_A-type_04 | 12 | TAATAAAAAAAT | 5'-POH |
| AT_x8_Network_T-type_01 | 12 | ATTTTTTATTAT | 5'-POH |
| AT_x8_Network_T-type_02 | 12 | ATATTTTTTATT | 5'-POH |
| AT_x8_Network_T-type_03 | 12 | ATTTTTTTTATT | 5'-POH |
| AT_x8_Network_T-type_04 | 12 | ATTTTTTTATTA | 5'-POH |

## 21. Supplementary Information Sources